\documentclass[aip,reprint, amsmath,amssymb,author-year]{revtex4-1}

\usepackage{graphicx}

\draft 

\begin{document}

\newcommand{\nc}{\newcommand}
\nc{\oln}{\overline}
\nc{\alphak}{\alpha^k}
\nc{\alphad}{\alpha^d}
\nc{\alphahat}{\hat\alpha}
\nc{\alphastar}{{\alpha^*}}
\nc{\aprime}{a'}
\nc{\bi}{\begin{itemize}}
\nc{\ei}{\end{itemize}}
\nc{\betahat}{\hat\beta}
\nc{\betadiss}{\beta^\dagger}
\nc{\betap}{\beta_p}
\nc{\be}{\begin{equation}}
\nc{\ee}{\end{equation}}
\nc{\bea}{\begin{eqnarray}}
\nc{\eea}{\end{eqnarray}}
\nc{\betad}{{\oln\beta_d}}
\nc{\betapart}{\beta_p}
\nc{\betapartd}{\beta_p^d}
\nc{\betastar}{{\beta^*}}
\nc{\betastard}{{\beta^{*,d}}}
\nc{\betaomegadplus}{\hat\beta_d}
\nc{\bexp}{b}
\nc{\bfI}{{\bf I}}
\nc{\bfM}{{\bf M}}
\nc{\bfr}{{\bf r}}
\nc{\bfT}{{\bf T}}
\nc{\bftau}{{\bf \tau}}
\nc{\bfig}{\begin{figure}}
\nc{\efig}{\end{figure}}
\nc{\bigR}{{\cal R}}
\nc{\bfA}{{\bf A}}
\nc{\bfB}{\bf B}
\nc{\bfD}{\bf D}
\nc{\bfJ}{{\bf J}}
\nc{\bfR}{{\bf R}}
\nc{\bfu}{{\bf u}}
\nc{\bfU}{{\bf U}}
\nc{\bfv}{{\bf v}}
\nc{\bfvi}{{\bf v}^{int}}
\nc{\bfV}{{\bf V}}
\nc{\bfx}{\bf x}
\nc{\calS}{{\cal S}}
\nc{\calSd}{{\cal S}^d}
\nc{\calSdplus}{\hat{\cal S}^d}
\nc{\calSdag}{{\cal S}^\dagger}
\nc{\calSolaf}{{\cal S^\ddagger}}
\nc{\calSwilcox}{{\cal S}}
\nc{\calSdwilcox}{{\cal S}^d}
\nc{\Cd}{C_D}
\nc{\dt}{\Delta t}
\nc{\diss}{\epsilon}
\nc{\dissd}{{\epsilon^d}}
\nc{\ensemble}{rms}
\nc{\ep}{a}
\nc{\epd}{{a^d}}
\nc{\epbar}{{\oln \ep}}
\nc{\epdbar}{\oln{\ep}^d}
\nc{\epsdav}{\overline{\ep}^d}
\nc{\epprime}{{\ep}'}
\nc{\Fcoef}{K_F}
\nc{\Fcoefd}{K_F^d}
\nc{\gammahat}{\hat\gamma}
\nc{\lan}{\langle}
\nc{\ran}{\rangle}
\nc{\lft}{\left}
\nc{\rgt}{\right}
\nc{\lab}{\label}
\nc{\f}{\frac}
\nc{\fdag}{f^\dagger}
\nc{\Fdag}{F^\dagger}
\nc{\inti}{\int_{-\infty}^{+\infty}}
\nc{\ob}{\overbrace}
\nc{\kd}{k^d}
\nc{\kdag}{k}
\nc{\kolaf}{k}
\nc{\kolafd}{k^d}
\nc{\kwave}{k_\lambda}
\nc{\kwilcox}{k}
\nc{\ult}{\widetilde u_l}
\nc{\lpp}{u_l^{''}}
\nc{\mud}{\mu^d}
\nc{\mudragtot}{\mu^{\ddagger}}
\nc{\mudragtotd}{\mu^{\ddagger,d}}
\nc{\mudragtotplus}{\hat\mu^{\ddagger}}
\nc{\mudragtotdplus}{\hat\mu^{\ddagger,d}}
\nc{\mutotplus}{{\hat\mu}^{tot}}
\nc{\mutotdplus}{{\hat\mu}^{tot,d}}

\nc{\muturb}{\mu_T}
\nc{\muturbd}{\mu_T}
\nc{\muturbplus}{{\hat\mu_T}}
\nc{\muturbdplus}{{\hat\mu_T^d}}
\nc{\muturbdagger}{\mu_T^\dagger}
\nc{\muturbdaggerd}{\mu_T^{\dagger,d}}

\nc{\mubar}{\oln{\mu}}
\nc{\mubardrag}{{\oln{\mu}^{drag}}}
\nc{\mubardAM}{\oln{\mu}^{AM}}
\nc{\mubardAMplus}{\oln{\hat\mu}^{AM}}

\nc{\omegad}{{\omega^d}}
\nc{\omegaplus}{\hat\omega}
\nc{\omegadplus}{{\hat\omega^d}}
\nc{\omegadag}{{\omega^\dagger}}
\nc{\omegaolaf}{{\omega^\dagger}}
\nc{\omegaolafd}{{\omega^{d,\dagger}}}
\nc{\omegawilcox}{{\omega}}
\nc{\omegawilcoxd}{{\omega^d}}
\nc{\phislowk}{\{\phi^k\}}
\nc{\phifastk}{{\phi^k}'}
\nc{\pprime}{p'}
\nc{\pbar}{\oln{p}}
\nc{\pfast}{\delta p}
\nc{\pslow}{\{p\}}
\nc{\phat}{\oln p}
\nc{\uiprime}{u_i'}
\nc{\uit}{\widetilde u_i}
\nc{\uipp}{{u_i^{''}}}
\nc{\ujprime}{u_j'}
\nc{\ujt}{\widetilde u_j}
\nc{\ujpp}{{u_j^{''}}}
\nc{\uippbar}{\oln{u_i^{''}}}
\nc{\ujppbar}{\oln{u_j^{''}}}
\nc{\vibar}{\oln v_i}
\nc{\vjbar}{\oln v_j}
\nc{\vlbar}{\oln v_l}
\nc{\vit}{{\widetilde v_i}}
\nc{\vipp}{v_i^{''}}
\nc{\vjt}{\widetilde v_j}
\nc{\vjpp}{v_j^{''}}
\nc{\vlt}{\widetilde v_l}
\nc{\vlpp}{v_l^{''}}
\nc{\vypp}{v_y^{''}}
\nc{\vip}{v_i'}
\nc{\vjp}{v_j'}
\nc{\vlp}{v_l'}
\nc{\non}{\nonumber}
\nc{\p}{\partial}
\nc{\pp}{p'}
\nc{\rhoc}{\rho^c}
\nc{\rhod}{\rho^d}
\nc{\rhoratio}{\lft(\f{\rhod}{\rho}\rgt)}
\nc{\rhobar}{\oln{\rho}}
\nc{\Rer}{R_{p}}
\nc{\Rp}{R_p}
\nc{\sigmahat}{\hat\sigma}

\nc{\sigmad}{\sigma^d}
\nc{\sigmadag}{\sigma^\dagger}
\nc{\sigmaolaf}{\sigma^\ddagger}
\nc{\sigmawilcox}{\sigma}
\nc{\sigmawilcoxd}{\sigma^d}
\nc{\sigstar}{{\sigma^*}}
\nc{\sigmastar}{\sigstar}
\nc{\tauint}{\tau}
\nc{\taurel}{\tau_p}
\nc{\taupart}{\tau_p}
\nc{\taupp}{\tau^{''}}
\nc{\tauslow}{\{\tau\}}
\nc{\ulpp}{u_l^{''}}
\nc{\uslow}{\{u\}}
\nc{\uslowi}{\{u_i\}}
\nc{\uslowj}{\{u_j\}}
\nc{\uxpp}{u_x^{''}}
\nc{\uxppbar}{\oln{u_x^{''}}}
\nc{\uxp}{{u_x'}}
\nc{\uyp}{u_y'}
\nc{\uypp}{u_y^{''}}
\nc{\uyppbar}{\oln{u_y^{''}}}
\nc{\uxt}{\wt U}
\nc{\vxt}{\wt U^d}
\nc{\vxtplus}{\wt{\hat U}^d}
\nc{\vxpp}{v_x^{''}}
\nc{\viprime}{v_i'}
\nc{\vjprime}{v_j'}
\nc{\Xchark}{X^k}
\nc{\yt}{\widetilde{U_y}}
\nc{\vyt}{\widetilde{V_y}}
\nc{\ypp}{u_y^{''}}
\nc{\wt}{\widetilde}
\nc{\beqs}{\begin{equation}}
\nc{\eeqs}{\end{equation}}
\nc{\beq}{\begin{equation}}
\nc{\eeq}{\end{equation}}
\nc{\beqa}{\begin{eqnarray}}
\nc{\eeqa}{\end{eqnarray}}
\nc{\lesfil}[1]{\input{#1}}

\nc{\deltanu}{\delta_\nu}
\nc{\utau}{u_\tau}
\nc{\Retau}{R_{e,\tau}}
\nc{\tunit}{T}
\nc{\tauwall}{\tau_w}
\nc{\xplus}{{\hat x}}
\nc{\yplus}{{\hat y}}
\nc{\yoh}{y^h}
\nc{\kplus}{\hat k}
\nc{\kdplus}{\hat k^d}
\nc{\Ud}{U^d}
\nc{\Uplus}{{\hat U}}
\nc{\Udplus}{{\hat U^d}}
\nc{\calSplus}{\hat{\cal S}}

\nc{\uxppbarplus}{\oln{\hat{u}_x^{''}}}
\nc{\uyppbarplus}{\oln{\hat{u}_y^{''}}}
\nc{\sigmaplus}{\hat{\sigma}}
\nc{\sigmadplus}{\hat{\sigma}^d} 

\title{ A generalized  $k-\epsilon$ model for turbulence modulation in fluid-particle flows }  



\author{Roar Skartlien}
\email[]{roar.skartlien@ife.no}
\affiliation{Institute for Energy Technology, PO Box 40, NO-2027 Kjeller, Norway}

\author{Teresa L. Palmer}
\affiliation{Venabo Analytics, Kirkeveien 96, 1900 Fetsund, Norway}

\author{Olaf Skjæraasen}
\affiliation{Institute for Energy Technology, PO Box 40, NO-2027 Kjeller, Norway}


\date{\today}

\begin{abstract}
A  large amount of published data show that particles with  diameter above 10\% of the turbulence integral length scale ($D/l >0.1$) tend to increase the turbulent kinetic energy of the carrier fluid above the single-phase value, and smaller particles tend to suppress it.
A revised phenomenological model of the  $k-\epsilon$ type was developed to reproduce these effects with the correct asymptotic limit of no turbulence modulation for small particles, and augmentation for larger diameter solids. Particle-kinetic theory was used to derive the work exchanged between the particles and the fluid due to both drag and added mass forces to accomodate any particle/fluid density ratios including bubbles, droplets and heavy solids.  For the larger particles, we devised a new model for vortex shedding induced by the slip between the particles and the turbulent flow, due to particle inertia.  
Simple approximate formulae for the turbulence modulation were obtained through asymptotic analysis, for the purpose of application. 


The overall effect for solid particles is that augmentation for large diameter solids is due to vortex shedding,  and turbulence suppression for small diameters is  due to  mainly to turbulent drag forces and extra fluid dissipation.  The transition from suppression to augmentation around $D/l  = 0.1$ is  a robust feature  for a wide range of particle Reynolds  and Stokes numbers, but we could not prove this to be a general relation on a theoretical basis. Indeed, bubbles and droplets may not display turbulence augmentation at all  for the larger diameters due to moderate turbulence  levels needed  to prevent  breakup, and  the  velocity difference between particles and fluid may therefore be too low for vortex shedding to occur. On the basis of the model we find that some  data  for solids in vertical gas flow show very large turbulence augmentation that can only be due to gravitational settling rather than slip that is generated by the turbulent flow.
 
\end{abstract}


\maketitle 



\section{Introduction}

Particles with a   diameter above 10\% of the turbulence integral length scale ($D/l >0.1$)  tend to increase  the turbulence kinetic energy,
while smaller particles tend to suppress it. This is observed for  a large amount of data as reviewed in the classical paper by \cite{GORE1989}, and more recently by e.g., \cite{TanakaEaton2008} and \cite{Saber2015}. It is remarkable that this behavior
is found in very different flow settings with combinations of liquid-particle, gas-particle, horizontal flow, and vertical flow. 
Mainly PIV (particle image velocimetry) and LDA/PDA (Laser Doppler Anemometry) have been used to gather turbulence modulation data.

A long-sought-after goal has been  to predict  turbulence modulation in the presence of particles in terms of a few  dimensionless numbers. 
The challenge has been that turbulence  modulation  is controlled by several different physical processes, and it has been difficult to establish a  sufficiently general turbulence model.  It is still hope  that the research community will obtain "order from the mist of turbulent data points"     \citep{Lohse2008}. 
Altered production of turbulence including modified velocity gradients and vortex shedding from the particles, to the modification of turbulence dissipation on small scales are all important ingredients. Hence, the particle Stokes number, Reynolds number,   particle/fluid density ratio and particle volume fraction are all essential parameters, but neither one  of these parameters alone is sufficient to predict  the degree of turbulence modulation  \citep[e.g.,][]{GAI2020}.

A recent promising attempt  was made by  \cite{TanakaEaton2008} that introduced a particle momentum number $Pa = St Re_L^2 (\eta/L)^3$   to  predict whether we have turbulence  augmentation or attenuation.   $St$ is the Stokes number, $Re_L$ is the flow Reynolds number, $L$ is the length scale of the flow, and $\eta$ in the Kolmogorov scale. This relation   was derived from the Navier-Stokes equation with  coupling   to the particles via the drag force only. 
Thirty independent data-sets for flow Reynolds numbers around $10^4$ were analyzed.  For $Pa <10^3$  the turbulence was augmented, for $10^3 < Pa < 10^5$  it was attenuated, and, surprisingly,  for $Pa > 10^5$ it was  augmented again. Although promising, a  model to predict the magnitude of the turbulence modulation was not offered.

\subsection{Modelling strategies}

 With the current work, we develop a new phenomenological model  where all the relevant non-dimensional numbers are incorporated.   We did  not adopt a predefined set of non-dimensional numbers, but rather built the model from the turbulence kinetic energy equation (k-equation).    
The k-equation can be derived from a suitable volume and ensemble average of the Navier-Stokes equation with the particle forcing included.  It suggests only two basic mechanisms: 1) production/loss  due to interaction  between the particles and the fluid,  2) modified production due to a possible change in the mean velocity gradient.  Two other effects do not emerge automatically from such an approach: 
 3) increased  dissipation due to the introduction of inter-particle length scales that enhances the small scale velocity gradients and 4) vortex shedding that injects vorticity on the particle diameter scale and smaller.

\cite{KENNING1997} and \cite{CROWE2000} emphasized that   particles  enhance   dissipation through increased shear rate  between particle pairs, and formulated a model for a dissipation length scale $l_h$ that decreases with smaller   particle separation and limits to the single-fluid length scale for zero volume fraction. It was assumed that the dissipation rate  scales as $k^{3/2}/l_h$.   However, if the  volume fraction is held constant  and the particle diameter is reduced, the dissipation would increase without limit, as $l_h$ approaches zero.  We  modified Kenning and Crowes model to remedy this problem. 

\cite{HETSRONI1989} discussed the importance of vortex shedding as a  source of turbulence augmentation, based on  data for the larger diameter particles ($D>0.1l$).
To account for vortex shedding, we developed a new modelling approach to account for slip between the particles and the turbulent fluid that occurs due to particle inertia ($St >0$). We emphasize that this approach accounts for the intrinsic effect of slip due to turbulence. Earlier work often treated the vortex shedding source with a prescribed characteristic slip velocity, in many cases set equal to the settling velocity for vertical flow.    

A number of DNS studies of 2-way coupling have been carried out by tracking the particles in the fluid with a prescribed equation of motion, and with back-reaction on the fluid by averaging the particle forces over the grid volume (e.g., \cite{AHMED2000, FERRANTE2003}). 
To incorporate vortex shedding and enhanced inter-particle velocity gradients, one would need to resolve the boundary layer of the particles, and this would soon be computationally prohibitive.
Therefore, Reynolds-averaged turbulence models have been used  to a large extent,  and they have almost exclusively been  based on the k-$\epsilon$ framework (e.g., \cite{Elgobash1983, KATAOKA1989,  Wang1997}). 
A variety of phenomenological modelling efforts have been reviewed recently by \cite{GAI2020}.
  A common factor has been to use approximate drag-models for the fluid-particle interaction terms, and   the added mass effect and other forces have been  ignored, thus restricting the model to high density ratio solids. 
  
The  added mass force is important for bubbles and droplets in liquid where the particle mass density is smaller or comparable to that of the carrier fluid.  The respective interaction terms emerge  naturally in the k-equation by considering the particle equation of motion (EOM) with the drag and added mass forces   retained, and when considering appropriate volume   and ensemble averaging \citep{2way2015}. By exploiting  kinetic theory for particles in turbulence \citep{REEKS92}, \cite{SKARTLIEN2009} showed that the work terms    can be recast into relatively simple algebraic formulae in terms of  correlation times associated to the cross and auto-correlation functions of the fluctuating forces between the fluid and the particles. If these closure parameters are not know a-priori, they have at least a clear physical interpretation, and they can   be measured using PIV \citep{SKARTLIEN2009}. 



\subsection{Objectives and scope}
The goal was to develop a simplified algebraic model for the turbulent kinetic energy in bulk regions, or core regions of turbulent flow consisting of a continuous phase with dispersions (bubbles, drops or solids). The context for application could be a layered flow with dispersions, occupying a fraction of a pipe diameter. We ignore boundary layer effects and assume that the mean velocities are given at at the boundaries of the layer, and ignore turbulent diffusion since the gradient of the turbulent kinetic energy in the core region is relatively small.   To this end, we assume that the volume fraction of the dispersion is a given constant, and hence the model is one-way coupled. It would be relatively straightforward to couple the model to a dispersion profile model for a full two-way model,  where the turbulence level from the model can drive the dispersion model.

\section{The k-equation, including drag and added mass forcing} 

 
\subsection{The particle equation of motion}

\cite{MAXEY1983} derived the equation of motion (EOM) for a particle
in a fluid in the limit of low particle Reynolds number.
We will neglect history   and   Faxen curvature effects,
but maintain the added mass contribution in terms of the difference between fluid and particle acceleration, and account for viscous stress and pressure gradients in the fluid, acting on the particle surface. The resulting EOM is 
\beq
\dot{\mathbf{v}}=
\frac{1}{\tau_p}(\mathbf{u}-\mathbf{v})
+\mathbf{g}_e+\alpha\frac{D\mathbf{u}}{Dt},
\label{eq:eom_simple}
\eeq
with the following parameters  
\beqa
\tau_p&=&\hat{\tau}_p(\mathbf{u},\mathbf{v})(1+\frac{1}{2}\frac{\rho_f}{\rho_p}) \nonumber\\
\alpha&=&\frac{\frac{3}{2}\frac{\rho_f}{\rho_p}}{1+\frac{1}{2}\frac{\rho_f}{\rho_p}} \nonumber \\
\mathbf{g}_e&=&\mathbf{g}
\left( \frac{1-\frac{\rho_f}{\rho_p}}{1+\frac{1}{2}\frac{\rho_f}{\rho_p}} \right), \nonumber 
\eeqa
where ${\rho_f}/{\rho_p}$ is the fluid/particle material density ratio. It is important to note that the fluid velocity is meant to be the undisturbed velocity without a particle present or the velocity of the ambient flow at sufficient distance from the particle.

In the limit of passive tracers, small particles follow the fluid, and the drag term  vanishes. However,   both the relaxation time $\tau_p$  and the velocity difference vanishes, posing a delicate "0/0" limit. The particle acceleration must now be identical to the fluid acceleration and the added mass contribution now vanishes (the difference between the fluid and particle acceleration is zero), but viscous stress and pressure gradients in the fluid acting on the particle surface provides the needed fluid acceleration term that balances the particle acceleration.
The primary added mass parameter is $\alpha$, which is in the range $[0,3]$ from high density to low density particles (solids to bubbles). To balance the EOM in the limit, we conclude that    
the particle density  must be equal to the fluid density such that $\alpha = 1$, and this also conforms with   zero buouancy for the passive tracer.

The particle relaxation time $\hat{\tau}_p(\mathbf{u},\mathbf{v})$  depends on the
local drag coefficient and the local particle and fluid velocities.  For small particle Reynolds number one recovers the Stokes time
\beq
\hat{\tau}_p=(\tau_p)_s=(\rho_p/\rho_f) D^2/(18\nu)
\eeq
where $\nu$ is the kinematic viscosity of the carrier fluid and $D$ is the particle diameter.  The relaxation time  is then
\beq
\tau_p=(\tau_p)_s[1+\frac{1}{2}\frac{\rho_f}{\rho_p}]= D^2[(\rho_p/\rho_f)+1/2]/(18\nu),
\label{eq:tauP}
\eeq
where the density correction of one half the density ratio is due to 
the viscous stress and pressure gradients in the fluid acting on the particle surface.
For bubbles and droplets, one must assume that the interfacial tension is high enough  so that a near spherical shape is maintained for the Stokes formula to be valid. A   particle Reynolds number correction was also incorporated. 

In kinetic theory, the  Stokes number  is defined by 
\beq
St=\frac{\tau_p}{\tau}, 
\eeq
where  $\tau$ is the auto-correlation time of the drag force seen by the particle.  To close the model, we adopt the approximation $\tau = \epsilon^{-1/3} D^{2/3}$, which is the turnover time of the turbulence at a length scale equal to that of the particle diameter. 

\subsection{The $k$-equation}

The turbulent kinetic energy of the fluid-particle mix in stationary channel flow can be expressed by   \citep{2way2015} 
\beqa
 \rho_f\epbar\calSwilcox_{xy}\partial_y \uxt
-\rho_f \epbar \epsilon + 
\p_y\lft[\lft(\mu + {\muturb}^{}\rgt)
\epbar\p_y k\rgt] + \non \\
\oln{F_i\uipp } + \rho_f W_v&=& 0,
\lab{eq:phasic1D_k1}
\eea
where the terms have dimension of energy rate per unit volume, $J/s/m^3$. An important aspect with this formulation is that the fluid velocity is   averaged over a small volume $V$ that encompasses a large number of particles. The velocity fluctuations are then the fluctuations of the volume averaged  velocity relative to the ensemble average.  The chosen averaging approach for the latter  is the so called phase average (Appendix \ref{App:Favre}),  enabling the separation of volume fraction and velocity related quantities in the k-equation. The fluctuating fluid velocity $\uipp$ is, for each vector component, 
\beq
\uipp = u_i-\uit= u_i-\f{\oln{\ep  u_i}}{\epbar}, \non
\eeq
 where $\ep$ is the volume fraction of fluid in $V$ and  $u_i$ is the average velocity taken over the {\it fluid volume} in $V$, and overbar denotes the ensemble average. Similar relations hold for the average particle velocity $v_i$, in $V$.  The associated turbulence kinetic energy in (\ref{eq:phasic1D_k1}) is defined as
\beq
k= \f{1}{2}\f{\oln{\ep\uipp\uipp}}{{\epbar}}.
\eeq
The phase averaged shear stress  $\calSwilcox_{xy}$ is defined in a similar manner (Appendix \ref{App:Favre}).
It can be shown that the standard Reynolds averages is recovered for small fluctuations in the volume fraction (but in terms of the volume average $u_i$). 


The first term  in (\ref{eq:phasic1D_k1}) 
is production due to the mean fluid velocity gradient, and $\uxt$ is   the  phase averaged   velocity in the  flow direction, 
\bea
\uxt &\equiv & \f{\oln{\ep  u_x}}{\epbar}.
\eea
The second term is dissipation in the fluid, and this factor will be derived later.  The third term is due to transport/diffusion of turbulent kinetic energy where ${\muturb}^{}$ is the eddy viscosity and ${\mu}$ the fluid viscosity. The last term $\rho_f W_v$ is added in the current work, and represents turbulence injection by  vortex shedding in the wake of the particles. 

The fourth term in (\ref{eq:phasic1D_k1}) 
 is the    work performed on the fluid in the small averaging volume.
Summation is implied over all indices $i=x,y,z$. The total force on the fluid   per unit volume  is
\beq
\mathbf{F}=-\f{1}{V} \sum_j m_p \dot{\mathbf{v}}_j \equiv -\f{N}{V} m_p \left[   \dot{\mathbf{v}}_j \right] = -\epd \rho_p \left[\dot{\mathbf{v}}_j \right],
\eeq
where brackets denote  averaging over all particles $j$ in the  volume $V$.  This volume 
must be considerably  larger than the average particle separation.  Again, the acceleration for any particle $j$ is $\dot{\mathbf{v}}_j$ as given by the EOM.
We assume that the volume 
averaged fluid velocity $\mathbf{u}$  can be taken as the common ambient (or far-field) fluid velocity in the Maxey-Riley formalism.
The work term can be expressed as 
\begin{widetext}
\bea
\oln{F_i\uipp} &=& 
\rho_p\beta \lft[ (\vxt - \uxt)\uxppbar
+ \oln{\epd\vipp\uipp} -  \oln{\epd\uipp\uipp}\rgt] +\rho_p \alpha  \lft(\epdbar\calS_{xy}\p_y\uxt \rgt),
\lab{eq:ks}
\eea
\end{widetext}
where the drag  coefficient is $\beta =  1/\tau_p $, the ensemble averaged dispersed  volume fraction is $\epdbar=1-\epbar$, and  $\vxt$ is  the  phase averaged   velocity  for the dispersed  phase along the mean flow direction, 
\bea
\vxt &\equiv & \f{\oln{\epd u_x}}{\epdbar} .
\eea
 
The first  term in (\ref{eq:ks}) is due to the difference in the averaged velocities between the fluid and the particles.  
 The following relations hold true for the phase averaging we have used (Appendix \ref{App:Favre}),
\bea
\uxppbar & \equiv & -\f{\oln{a'\uxp}}{\oln{a}} = \f{\oln{(\epd)' \uxp}}{\oln{a}},
\label{eq:DEFufluct}
\eea
where $()'$ refers to the deviation from the normal ensemble average, and  
$a'+(\epd)'=0$. 
Hence, $\uxppbar$  can be interpreted as a turbulent volume flux in the axial direction.  
The third  term in (\ref{eq:ks})
represents removal of fluid kinetic energy due to the fluctuating part of the drag force, and is always negative. 
To close this term we assume sufficiently small volume fraction fluctuations so that 
\bea
\oln{\epd\uipp\uipp} \simeq  \f{\epdbar}{\epbar}\lft(\oln{\ep\uipp\uipp}\rgt) \simeq 2\epdbar k. \non\
\eea
The last term in (\ref{eq:ks}) is due to the correlation between the fluid velocity and the fluid acceleration forcing in the EOM. For passive tracers, the drag-related  terms vanish, and $\rho_p=\rho_f$ so that $\alpha=1$  and  the last term 
adds to the gradient production term in (\ref{eq:phasic1D_k1}). And since $\epdbar+\epbar=1$, the single phase form of the gradient production is recovered from (\ref{eq:phasic1D_k1}).

It is important to note that both (\ref{eq:phasic1D_k1}) and  (\ref{eq:ks}) are defined in terms of an average over a volume $V=L^3$. Larger volume corresponds to a more "severe" lowpass filtering of the turbulence energy spectrum, and the modelled $k$ would be reduced. The vortex shedding term in (\ref{eq:phasic1D_k1}) represents the smaller scales $\kappa > 2\pi/D$ above the filter cutoff at $\kappa \sim 2\pi/L$, and is not subject to the filter.
The  reduction of $k$  as function of $V$ is discussed  in Appendix \ref{App:Favre}, and it is argued that the reduction is  small  when $L$ is a few times the average particle separation $\lambda$. With this assumption, it is not necessary to add potential filter coefficients in  k-equation. These concepts are illustrated further in Figure \ref{fig:scales}.

\subsection{Closure for the fluid/particle velocity correlation function}

The second source term in (\ref{eq:ks}) can be positive and represents  the work done on the fluid by the fluctuating part of the drag  force.
To close this term using kinetic theory, we adopt the approximation
\bea
\oln{\epd\vipp\uipp} &\simeq& \epdbar \oln{ v'_i u'_i} 
\eea
where $(..)'$ is the fluctuation relative to the straight ensemble average, 
\bea
v'_i&=&v_i-\oln{v_i} \non \\
u'_i&=&u_i-\oln{u_i}. \non
\eea
This approximation is accurate for small fluctuations in the volume fraction, where one can put $\epd \simeq \epdbar$ and consequently $\vipp \simeq v'_i$  and $\uipp \simeq u'_i $ (Appendix \ref{App:Favre}). Again, it is noted that $v_i$ represents the volume averaged particle velocity and  $u_i$ the volume averaged fluid velocity.
The general closure relation is obtained via kinetic theory, and the result is  
\bea
\oln{ v'_i u'_i}  
= \tau_p \lft[ (\overline{ \lambda}_{ii})_{dd}/\tau+(\overline{ \lambda}_{ii})_{da}/\tau_{ad} \rgt],
\eea
using  the  dispersion tensor $\overline{\lambda}$ from the \cite{REEKS92} theory (Appendix \ref{app:lambda}). Here, $\tau_{ad}$ is the cross-correlation time between the drag force and the added mass force seen by the particle, and $\tau$ is the auto-correlation time of the drag force seen by the particle.    
The total work can be written in terms of an "average slip velocity source" $S_b$, and a "fluctuation source" $S_f$,    
\bea
\oln{F_x\uxpp} &=&
\rho\beta \lft[ S_b
+ \epdbar S_f \rgt] +\f{\rho}{2}\lft(\epdbar\calS_{xy}\p_y \uxt \rgt),
\lab{eq:stress_1Dcouple1}
\eea
 we obtain the following algebraic form for $S_f$,
 \begin{widetext}
\bea
 S_f&=&\lft[ \oln{\epd\vipp\uipp} -  \oln{\epd\uipp\uipp}\rgt]/\epdbar \simeq \lft[ \tau_p \left( \frac{(\overline{\lambda}_{ii})_{dd}}{\tau} +  \frac{(\overline{\lambda}_{ii})_{da}}{\tau_{ad}} \right) -  2k\rgt]    \\
&=& -2k\frac{St}{1+St}+\overline{u'_x u'_y}\p_y\uxt\left[ \frac{\tau}{(1+St)^2}+\frac{\alpha\tau_{ad}}{1+\beta\tau_{ad}} \right].    
\label{eq:SF1}
\eea
\end{widetext}
The first term in (\ref{eq:SF1})   is dominating for the cases we studied and the magnitude  increases with particle diameter as  $St \sim D^{4/3}$. The asymptote is $-2k$ due to the "$-uu$" term, while the "$vu$" term approaches zero.  The same term limits to zero for  small $St$ as the particles become passive tracers. 
The smaller second  term in (\ref{eq:SF1})  also originates from the "$vu$" term, and is a function of the velocity gradient normal to the mean flow direction (y-direction). Cross flow particle motion in the y-direction induces axial motion in the x-direction via drag and corresponding axial work against the fluid. This term also tends to zero for small particles  when $\tau$ and $\tau_{ad}$ diminish. 
Furthermore $\beta$ is now large  and removes the last term inside the brackets.   

In the limit of passive tracers, the work (\ref{eq:stress_1Dcouple1}) depends on
$\beta S_f \sim  p \beta St + q \beta \tau = p/\tau + q/St$ ($p$ and $q$ are constants).  A paradox is that this appears to diverge  for $St \rightarrow 0$   while there  should be no work exchange  for passive tracers.  
The source of the problem is that $S_f$ does not approach zero fast enough relative to the increasing value of the drag coefficient $\beta$. 
Hence, the model is not asymptotically correct for infinitesimal particle diameters.  
However, the problematic asymptote is avoided since 
the particle diameter should remain larger than the Kolmogorov scale  for direct turbulence modulation, so that the Stokes number cannot be arbitrarily small.  

\subsection{Closure for the average slip-velocity term}

The slip term is  
\bea
S_b=\lft[ (\vxt - \uxt)\uxppbar\rgt], 
\eea
and by invoking  (\ref{eq:DEFufluct}) we require a 
closure relation for the axial turbulent volume  flux of particles, $\oln{a'_d\uxp}$. It is fortunate that this is given in terms of the diffusion current of \cite{REEKS92},
\bea
\overline{a'_d u'_x}=-\tau_p [ \partial_{y}(\epdbar\overline{\lambda}_{yx}) + \epdbar \overline{\gamma}_{x}]_d,
\eea
where $[..]_d$ means inclusion of only the drag component of the  dispersion tensors.
Since there is no variation of the turbulence level in the axial direction, there is  zero axial drift velocity so that $\overline{\gamma}_{x}=0$.  The closure relation for $\uxppbar$ can now be reduced to the algebraic form (Appendix \ref{app:lambda})
\beq
\uxppbar \simeq - \beta   \lft[ \frac{\tau^2}{1+\beta\tau-\alpha C_{yy}^{''}\tau^2} \rgt]  \frac{\partial_{y}(\epdbar \overline{ u'_x u'_y})}{\epbar},
\label{eq:slip-closure}
\eeq
where $C_{yy}$ is the double derivative of the fluid normal stress in the y-direction. 
$S_b$ is positive if  the particles lead the fluid with $(\vxt - \uxt) > 0$, provided that the shear stress and dispersion gradient are such that $\uxppbar > 0$.  
As the model is constructed for bulk flow regions,   the normal stress curvature across layer can be neglected so that   $C_{yy}^" \simeq 0 $.
We found that $S_b$ had very little effect in all cases studied, within   reasonable magnitudes of the slip velocity $(\vxt - \uxt)$.

\section{Dissipation and production: a phenomenological approach}

The rigorous treatment
for the fluid-particle work term including drag and added mass is a necessary foundation, but not a sufficient step to reproduce the trends in the data. It was clear that consistency with the  data could only be achieved by adding extra dissipation  and the extra source due to vortex shedding. It is important to realize that these effects cannot  emerge naturally from  
averaging the Navier-Stokes or the k-equation over the appropriate fluid volume.  Volume averaging implies an effective lowpass filtering of the energy spectrum, and the associated fluid-particle work occurs on a "meso-scale", spanning multiple interparticle distances.   In contrast, vortex shedding and extra dissipation must both be   considered  "micro-scale" effects on length scales comparable to or smaller than  the particle diameter. Hence, the energy contained in the smaller scales must be put back into the k-equation to represent the total turbulence kinetic energy.

\begin{figure*} 
\includegraphics[width=0.6\linewidth]{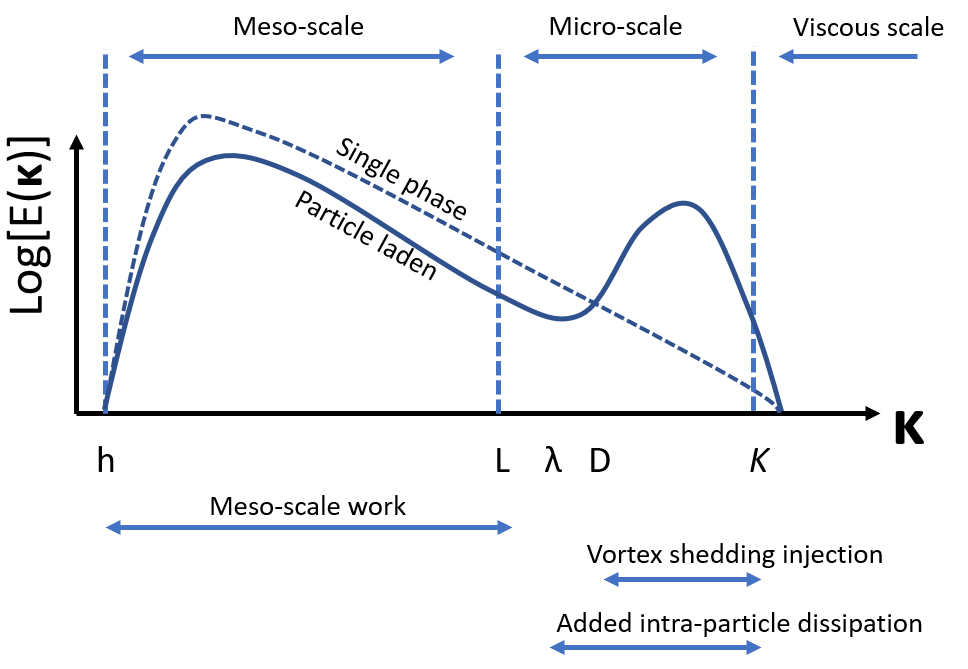}
\caption{\label{fig:scales} Schematic illustration of the scales involved in turbulence modulation and how they relate to the turbulence energy spectrum. The averaging volume is $L^3$, $\lambda$  is the mean particle separation, $D<\lambda<L$ is the (variable) particle diameter, and $K$ is the Kolmogorov scale. }
\end{figure*}

 These ideas are further illustrated in the "schematic"  Figure~\ref{fig:scales} in terms of the turbulence energy spectrum as function of wave number $\kappa$. The meso-scale range is from the flow scale $h$ (layer width) to the relatively small averaging length $L$ that defines the averaging volume. 
 $\lambda<L$  is the mean particle separation. $D<\lambda$ is the particle diameter, and $K$ is the Kolmogorov length. Extra dissipation occurs for scales smaller than $\lambda$, and turbulence injection by vortex shedding occurs in the micro-scale range between $D$ and $K$.  The meso-scale work is negative, implying a suppressed energy in the meso-scale range, while turbulence injection leads to increased energy in the micro-scale range due to vortex shedding. The energy spectrum of the particle laden flow is indicated together with the unladen, single-phase version. The total turbulent kinetic energy $k$ corresponds to the integral of the spectrum.  
 
 \subsection{Dissipation}

In situations of high Reynolds number and where the turbulence transport terms are negligible, the dissipation rate ($J/kg/s$) in particle-free flow can be modeled as \citep{LS72, wilcox2006turbulence} 
\bea
\epsilon 
= C_\mu \f{k^{3/2}}{l} \; \text{single-phase},
\eea
where  
$l$ is the characteristic integral length scale of the turbulence and  the empirical constant $C_\mu \simeq 0.09$ for single phase fluids.
We  assume that the turbulence length scale is a function of the Reynolds number of the layer and obeys the same type of scaling law as for pipe flow,
\beq
l=chRe^{-1/8},
\eeq
where $Re=h \Delta U /\nu_c$ set by the velocity difference over the layer we consider (of thickness $h$), rather than the bulk average velocity, and the factor $ 0 < c < 1$ so that $l\sim 0.1 h$ approximately. The kinematic viscosity of the continuous fluid is $\nu_c$. This definition of $l$ must be considered as a single phase value, as the Reynolds number above is independent of the turbulence modulation or the presence of the particles.
 
 To account for particles, \cite{KENNING1997} and \cite{CROWE2000}  assumed the modified form
\bea
\epsilon =  C_\mu \f{k^{3/2}}{ l_h} \; \text{with particles},
\label{eq:DissCrowe}
\eea
where $l_h< l $ is the  "hybrid" length scale
\bea
\f{1}{l_h}=\f{1}{\lambda}+\f{1}{l},
\eea
and where the mean particle separation  is 
\beq
\lambda= D \lft(\f{\pi}{6\epdbar} \rgt) ^{1/3}.
\eeq
 The harmonic mean gives more weight to the separation length scale if this is smaller than $l$, and dissipation will be enhanced. The physical interpretation is that enhanced local velocity gradients between the particles increases the viscous dissipation rate. 
 
\begin{figure}
\centering
\includegraphics[width=0.8\linewidth]{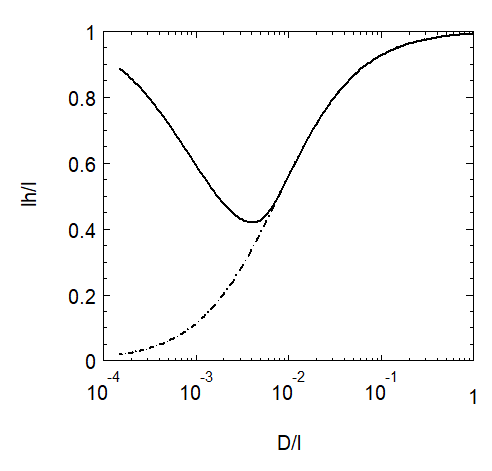}
\caption{Example of the hybrid length scale ratio $l_h/l$  with  the weight function $C_\lambda(D)$ with  $k_\lambda$ =0.05 and an average $D_0=24 \mu$m (full upper curve). The integral length scale was here $l=7.75$mm. The dashed line is without the weighting ($C_\lambda=1$) and this would give  vanishing dissipation length scale as the particle diameter approaches zero, and a corresponding infinite dissipation. Note that the weighted variant correctly limits to $l$ for both small and large $D$, as $l_h/l$ limits to 1.0. This example corresponds to  the case  of neutrally buoyant solids in water. }
\label{fig:lh}
\end{figure}

Unfortunately, this form did not reproduce the experimental evidence of relatively small or no turbulence suppression  for $D/l \rightarrow 0$.
The problem is that as the particle diameter goes to zero for a fixed volume fraction, the particle separation goes to zero, and the dissipation  would be unlimited with   100\% turbulence suppression.    The heart of the problem is that sufficiently small particles 
tend to follow the flow passively and they will be a part of the continuum of the fluid rather than objects that can modify the velocity field in the inertial sub-range of length scales. Very small passive tracer particles will not influence the  turbulent dissipation rate, except for enhancing the effective kinematic viscosity and density of the of the fluid. This is not captured with the hybrid length scale as it is presented above.
To remedy the problem,
we introduce a cutoff-weight $C_{\lambda}$, 
\bea
\f{1}{l_h}=\f{C_{\lambda}}{\lambda}+\f{1}{l}.
\eea
The weight goes to zero exponentially for small diameter such that the hybrid scale   limits to   $l$ as $\lambda \rightarrow 0$,
\bea
C_{\lambda}=k_\lambda (1-e^{-(D/D_0)^2}),
\eea 
where $D_0$ is taken to be proportional to the Kolmogorov scale  at which the turbulent fluctuations are no longer significant, and viscous dissipation takes over.  There will be an intermediate range of diameters where the hybrid scale is smaller than $l$ and where the dissipation is elevated due to the presence of particles (Figure \ref{fig:lh}) as intended by Crowe.
A normalization of the dissipation rate $\epsilon$ ($J/s/kg$) with the fluid volume fraction $\epbar$ was necessary to recover the unmodified turbulence level in the passive tracer limit, and we adopted 
\bea
\epsilon =   \frac{C_\mu}{\epbar} \f{k^{3/2}}{ l_h}.  
\eea
This implies that the dissipation rate {\it per unit mass of fluid} should increase when the  amount of fluid in a volume $V$ decreases when the particle volume fraction increases.

\subsection{Production by vortex shedding}

Production of turbulence by vortex shedding in the wake of particles is a potentially  important contributor to turbulence modification \citep{HETSRONI1989}.  
Internal circulation in a droplet or gas bubble implies high interfacial mobility (as opposed to a solid particle with no-slip conditions at the surface) and this 
can drastically alter the wake structure leading to delayed boundary layer separation and smaller 
wakes compared to solids \citep{Clift1978}. 
For simplicity, we will assume contaminated bubble and droplet interfaces and adopt the same drag coefficient as for a solid.    

Only a part of the work performed on the fluid from a single particle is converted to turbulent kinetic energy in the wake,  provided that the particle Reynolds number is well above the critical value of vortex shedding. 
The  work (in $J/s$ ) on the fluid from a single particle is $W_p=F U_s$, where the slip velocity $U_s$ 
is the difference between the particle and the fluid velocities, with the latter referring to the far-field, asymptotic
fluid velocity interpolated to the particle’s position. 
$F$ is the force on the fluid from the particle (and vice versa with opposite sign). In terms of the drag coefficient $C_D$,
\beq
F=\f{1}{2} \rho_f C_D \pi (D/2)^2  U_s^2,
\eeq
and the work  per unit time ($J/s$) per particle is 
\beq
W_p = \f{\pi}{8}\rho_f C_D D^2 U_s^3.
\eeq
The drag coefficient for a spherically intact particle with little interfacial mobility is 
\beq
C_D=24/Re_p  (1+0.15Re_p^{0.687}),
\eeq
where the particle Reynolds number is in terms of the slip velocity $U_s$.
The total production of turbulent kinetic energy  due to vortex shedding  per unit mass  of fluid ($J/s/kg$) is 
\beq
W_v = k_v C_0 f_{Re} \f{\pi}{8} n_p C_d D^2 U_s^3,
\label{eq:vortexshed}
\eeq
where $n_p=6 \epdbar /\pi/D^3$ is the number density of particles.
The   scaling factors $C_0$ and $ f_{Re} $ are given in Appendix \ref{App:VSF}.

There will be two types of vortex shedding sources. One general "intrinsic" source due to particle inertia (non-zero  $St$) and the consequential lagging of the particles behind the turbulent motion of the flow. The other possible source is gravitational settling.
This can be labelled an "extrinsic" source since the forcing is now due to an external body force. Both forms were considered when comparing to experimental data.

\subsubsection{Intrinsic vortex shedding: Inertial slip}

For large Stokes number the particle responds "slowly" to the flow and  the particle velocity is relatively  small compared to the fluid velocity. Hence, the magnitude of the fluctuating slip velocity is large and comparable to that of the fluid velocity.  
For small Stokes number, the particle nearly follows the flow 
and the slip velocity is reduced. A particle of a specific diameter $D$  responds differently to the various length scales in the turbulence, where larger length scales have  larger  eddy turnover time, and hence the Stokes number is smaller.  The Stokes number   
varies with the eddy  turnover time  as  $St(s)=\tau_p(D)/\tau(s)$ where $s$ is the length scale variable, the eddy turnover time is $\tau(s)=\epsilon^{-1/3} s^{2/3}$ , and $\tau_p=bD^2$ is the relaxation time for a particle of diameter $D$, 
\beqa
\tau_p &=& D^2[(\rho_p/\rho_f)+1/2]/(18\nu)/(1+0.15 Re_p^{0.687}) \non \\ &\equiv& b D^2,
\eeqa
where $b$ has dimensions ($s/m^2)$.
We postulate that the total slip velocity is the integral over the turbulence energy spectrum, scaled with  a Stokes number dependent filter $W$ that varies over the spectrum due to the variation of the eddy turnover time,
\beq
W(s)=\left[ \frac{St(s)}{1+St(s)} \right]^n,
\label{eq:W}
\eeq
where $n$ is an exponent to be determined. Then for small $St$ (large scales $s$), the weight is small and the contribution to the slip velocity is small, and for large $St$ (small scales $s$), the contribution is larger.

The turbulent kinetic energy on scales smaller than the particle will probably not be able to induce a coherent boundary layer around the particle to generate efficient vortex shedding in the classical sense. Hence, we also  postulate that only  the turbulent kinetic energy on length scales equal to and larger than the particle diameter $D$ would be able to induce   vortex shedding.
The total filtered kinetic energy per unit mass  can then be written as the following integral over the turbulence spectrum 
in the inertial subrange,
\beq
\frac{U_s^2}{2} = C \epsilon^{2/3} \int_{\kappa_0}^{\kappa_D} W(\kappa) \kappa^{-5/3} d\kappa,
\eeq
where the factor $C \simeq 1.5 $,  $\kappa=2\pi/s$ , $\kappa_D=2\pi/D$, and $\kappa_0=2\pi/h$, where $h$ is the largest scale that contains turbulent energy (say the width of the channel, pipe or jet).  We  assumed the standard single-phase form of the Kolmogorov turbulence spectrum, as a first order approximation (neglecting turbulence modulation of the shape of the spectrum).
The filter function in terms of the wavenumber is
\beq
W(\kappa)=\left[ \frac{1}{1+1/St(\kappa)} \right]^n =  \frac{1}{ \left[ 1+(2\pi)^{2/3} \epsilon^{-1/3} \kappa^{-2/3}/\tau_p \right]^n }. \non
\eeq
With this form, we were fortunate to find the closed analytic form  
\begin{widetext}
\beq
U_s^2 = 2C\epsilon^{2/3} \int_{\kappa_0}^{\kappa_D}\frac{ \kappa^{-5/3}}{\left[1+(2\pi)^{2/3} \epsilon^{-1/3} \kappa^{-2/3}/\tau_p\right]^n}  d\kappa 
= 6C \frac{\epsilon^{2/3}}{2r(n-1)} \bigg\rvert_{\kappa_0}^{\kappa_D} \frac{1}{\left[ 1+r\kappa^{-2/3} \right]^{n-1}},
\eeq 
\end{widetext}
where $r=(2\pi)^{2/3}\epsilon^{-1/3}/(b D^2)$.
Increasing the filter exponent $n$ tends to select parts of the spectrum closer to  the particle diameter ($\kappa_D$), provided that $n>5/2$. Lower $n$ tends to select the lower wavenumbers closer to $\kappa_0$. 

The total  turbulent kinetic energy input from   vortex shedding (\ref{eq:vortexshed})  can now be written as  
\begin{widetext}
\beqa
W_v =  Q \left[ \frac{1}{2r(n-1)}  \right]^{3/2}  \left[\frac{1}{\left[ 1+r\kappa_D^{-2/3} \right]^{n-1}} -
\frac{1}{\left[ 1+r\kappa_0^{-2/3} \right]^{n-1}} \right]^{3/2}  (C_{\mu}/(\epbar l_h)) k^{3/2}, 
\eeqa
\end{widetext}
where 
\beqa
Q=k_v(6C)^{3/2} C_0 f_{Re} \f{\pi}{8} n_p C_D D^2.
\eeqa
It is the factor $f_{Re}$ that "switches on" the vortex shedding at the critical particle Reynolds number $Re_c$. We found that the magnitude of $W_v$ increases with diameter until a maximum value    occurs in the range $D/l \in [0.1, 1.0]$.
With the definition 
\beq
W_v = \epdbar B_V k^{3/2},
\eeq
we  define the vortex shedding coefficient    
\begin{widetext}
\beq
B_V=  k_v (6C)^{3/2}  C_{0} f_{Re} \f{3}{4D}  C_D  \left[ \frac{1}{2r(n-1)}  \right]^{3/2}  \left[\frac{1}{\left[ 1+r\kappa_D^{-2/3} \right]^{n-1}} -
\frac{1}{\left[ 1+r\kappa_0^{-2/3} \right]^{n-1}} \right]^{3/2}  (C_{\mu}/l_h).
\label{eq:BV}
\eeq
\end{widetext}
The dimension  of $B_V$ is inverse length, $m^{-1}$, as $r \sim D^{-2/3}$.  

\subsubsection{Extrinsic vortex shedding: External forcing}

The total production of turbulent kinetic energy  due to vortex shedding  per unit mass  of fluid ($J/s/kg$) is now 
\beq
W_{vg} = k_{v} C_0 f_{Re} \f{\pi}{8} n_p C_d D^2 U_{sg}^3,
\eeq
where  $U_{sg}$ is the slip velocity due to gravity or more generally, any external force acting on the particles. This can be taken as the terminal velocity under the appropriate experimental conditions, or from direct measurements of the average fluid and particle velocities.
If the extrinsic source is considered, then $W_{vg}$ is added to the intrinsic source $W_v$ in the model equations.

\section{A model for shear flow}

The model was tested in  horizontal sheared    flow   with either 
 gas bubbles, liquid droplets or solid particles. 
The average   velocity is assumed to be known at the upper and lower boundaries of the computational domain. By considering interior regions only (excluding turbulent boundary layer effects) one can neglect   cross-flow diffusion of turbulent kinetic energy as the gradient in the turbulent kinetic energy is small. One can  then 
formulate the problem in terms of a third order algebraic equation for  the turbulent kinetic energy.     

\subsection{Turbulent kinetic energy in the bulk flow}

Without diffusion, the  k-equation  reduces to 
\beqa
 \rho_f\epbar\calSwilcox_{xy}\partial_y \uxt
-\rho_f \epbar \epsilon 
+\oln{F_i\uipp } +\rho_f W_v &=& 0.
\lab{eq:phasic1D_k}
\eea
 The velocity gradient across  the layer can be estimated as 
\beq
\partial_y \uxt \simeq \Delta U / h,
\eeq
where $h$ is the layer thickness, and $\Delta U$ is the velocity difference over the layer. 
Furthermore, 
\beqa
-\calS_{xy}  &=& \f{\oln{\ep\uxpp\uypp}}{{\epbar}} \non  \simeq  \oln{\uxp\uyp} \simeq -\nu_T  \partial_y \uxt  \simeq  -\nu_T \Delta U / h \non \\ &\simeq& -\zeta  U^* \Delta U,  
\eeqa
where the kinematic eddy viscosity is $\nu_T$, and where the shear stress is replaced by the Boussinesq approximation  $\oln{\uxp\uyp}\simeq -\nu_T  \partial_y \uxt$.
The eddy viscosity scales as $ \nu_T \simeq \zeta h U^*$, with $\zeta=0.035$ and $U^*$ is the friction velocity. We adopt the characteristic value 
\beq
U^* \simeq \sqrt{2k}, 
\eeq 
and the  production can be expressed in terms of $\Delta U$ and $U^*$,
\beq
\calSwilcox_{xy}\partial_y \uxt \simeq   \nu_T (\Delta U / h)^2= \zeta  (U^*/h) (\Delta U)^2.
\eeq

The closure relation (\ref{eq:slip-closure})   contains 
\beq
\frac{\partial_{y}(\epdbar \overline{ u'_x u'_y} )}{\epbar}= 
\frac{\partial_{y}(\epdbar )}{\epbar} \overline{ u'_x u'_y} +
\epdbar\frac{\partial_{y}( \overline{ u'_x u'_y})}{\epbar}, 
\eeq
where we invoke the Boussinesq approximation once again.    
The gradient of the shear stress is  constant in pipe or channel flow with the value $2\tau_w/\rho/h = 2 (U^*)^2/h$ in terms of the wall shear stress.  We also assume  it is constant throughout the layer and use  
\beq
\partial_{y} \overline{ u'_x u'_y} \simeq  2 (U^*)^2/h.  
\eeq
The  average velocity slip contribution becomes 
\begin{widetext}
\beq
S_b\simeq (\uxt - \vxt) \tau_p \beta^2   \lft[ \frac{\tau^2}{1+\beta\tau-\alpha C_{yy}^{''}\tau^2} \rgt]  
\lft[ \frac{-\partial_{y}(\epdbar )}{\epbar} (\zeta U^*  \Delta U) + \frac{\epdbar}{\epbar} ( 2 (U^*)^2/h)  \rgt].
\label{eq:SB}
\eeq
\end{widetext}
We may ignore the first term in the second bracket if the gradient of the dispersed volume fraction is sufficiently small, and the slip work is positive if the particles lag the fluid on the average.
The drag related work term becomes  
\begin{widetext}
\bea
 S_f
&\simeq&  -2k\frac{St}{1+St}-(\zeta  U^* \Delta U) (\Delta U/h) \left[ \frac{\tau}{(1+St)^2}+\frac{\alpha\tau_{ad}}{1+\beta\tau_{ad}} \right] <0, 
\label{eq:SF}
\eea
\end{widetext}
and this is always negative leading to turbulence suppression. The first term is the dominating factor for the cases we studied.

The k-equation  per unit mass ($J/s/kg$) is  
\beqa
\epbar\calSwilcox_{xy}\partial_y \uxt
- \epbar \epsilon  
+\oln{F_i\uipp }/\rho_f + W_v &=& 0,
\lab{eq:knorm}
\eea
and this can now be formulated as a third order algebraic equation with the approximations developed above,  
\begin{widetext}
\bea
\zeta  \lft[ \epbar + \epdbar \alpha \f{\rho_p}{\rho_f}  \rgt]  (\sqrt{2}  /h) (\Delta U)^2 \xi -( C_\mu/l_h - \epdbar B_V) \xi^{3}  +C_0\beta \f{\rho_p}{\rho_f}  \lft[ 
S_b(\xi^2,\xi)  + \epdbar  S_f(\xi^2,\xi)  \rgt] =0,
\label{eq:master}
\eea
\end{widetext}
where $\xi=\sqrt{k}$, and $U^*= \sqrt{2}\xi$.
The first order term incorporates the gradient production. The third order term incorporates dissipation and production by vortex shedding. The last term is the work exchanged between the particles and the fluid.  It is convenient to split the latter contributions in first and second order terms,  
\beqa
S_b(\xi^2,\xi) = S_b^1(\xi)\xi + S_b^2(\xi)\xi^2  \non \\
S_f(\xi^2,\xi) = S_f^1(\xi)\xi + S_f^2(\xi)\xi^2 \non 
\lab{eq:masterB}
\eeqa
The particle Reynolds number and the Stokes number depends on the turbulence kinetic energy and turnover timescales  respectively, and the coefficients are therefore higher order algebraic functions of $\xi$. 
 An iterative solution of the third order equation is presented in Appendix \ref{App:iterate}.

 For zero volume fraction, the particle separation  $\lambda \rightarrow \infty$, and $l_h \rightarrow l$  and production will now  equal dissipation in the pure fluid. The unmodified level of turbulent kinetic energy is then set by
\bea
\xi^{2} = k_0 =    \zeta    (\sqrt{2} ) (\Delta U)^2 /( h C_{\mu}/l). 
\eea

\subsection{The limit of zero meso-scale work} 
\label{sec:zerowork}
The meso-scale work $\oln{F_i\uipp}$  was negative in  the cases we studied, and relatively small compared to production and dissipation.  Thus, a "maximum envelope" of the turbulent kinetic energy as function of particle diameter could be obtained by setting the meso-scale  work to zero.   
The envelope is given in terms of the balance between   dissipation, energy input from vortex shedding, and turbulence production due to the large scale velocity shear,  
\beqa
k_{max} =
 \frac{ \zeta  \lft[ \epbar + \epdbar \alpha \f{\rho_p}{\rho_f}\rgt]  (\sqrt{2}) (\Delta U)^2 } { hC_\mu \left(    \lft[ \f{C_{\lambda}}{\lambda}+\f{1}{l} \rgt] - \epdbar B_V/C_\mu \right)  },
\eeqa
and relative to the unmodified turbulence,
\beq
\frac{k_{max}}{k_0}   
  =   \frac{1- \epdbar (1-\alpha \f{\rho_p}{\rho_f}) }{  l/l_h - \epdbar  B_V l/C_\mu}.
  \label{eq:envelope}
\eeq
This potentially allows for   large turbulence augmentation through vortex shedding when $B_V l_h/C_\mu \rightarrow 1$. For   particles in the size range $D/l \in [0.1,1]$, \cite{GORE1989} reported  augmentation of   large  magnitude up of the order of  100\% or more, depending on the flow setting. 
For vanishing volume fraction ($\epdbar \rightarrow 0$), $\lambda \rightarrow \infty$ and $k_{max}= k = k_0$.

For relatively small particles or bubbles without any vortex shedding such that  $B_V \simeq 0$,
\beq
\frac{k_{max}}{k_0} \simeq  
  \frac{l_h}{l} \left[ 1- \epdbar (1-\alpha \f{\rho_p}{\rho_f}) \right]. 
  \label{eq:envelope2}
\eeq
For very small particle diameters that represent the tracer limit,  $C_\lambda/\lambda \rightarrow 0$  and
 \beq
\frac{k_{max}}{k_0} \simeq  1- \epdbar (1-\alpha \f{\rho_p}{\rho_f}),  
\eeq
and the modification is essentially controlled by the   added mass effect via  the $\alpha$ parameter.  
 For passive tracers $\alpha {\rho_p}/{\rho_f} = 1$ and $k_{max}=k_0$.

 \subsection{The limit of zero vortex shedding}
 \label{sec:zeroshed}
 Vortex shedding will be negligible for sufficiently small particle Reynolds number $Re_p$. For larger gas bubbles to stay intact, the turbulent kinetic energy has to be moderate and the slip velocity (and  $Re_p$) may then be too small to induce significant vortex shedding. 
With meso-scale work included, 
 \beq
 k = \xi^2 \simeq  \f{A+C_1+C_2\xi}{C_\mu/l_h}, 
 \eeq
where the work terms $C_1$ and $C_2$ are defined in Appendix \ref{App:iterate}, Equation (\ref{eq:C1}, \ref{eq:C2}).   The k-ratio is now approximately
 \beq
 \f{k}{k_0} \simeq  \f{l_h}{l} \lft[  1- \epdbar (1-\alpha \f{\rho_p}{\rho_f})  \rgt] +(C_1+C_2\sqrt{k})\f{l_h}{C_\mu k_0}.
 \label{eq:noshed}
 \eeq

 For bubbles, the terms $C_1$ and $C_2$ are negligible due to the low particle/fluid density ratio, and the added mass term in $\alpha$ can also be ignored for the same reason, and
 \beq
 \f{k}{k_0} \simeq  \f{l_h}{l} \lft[  1- \epdbar )  \rgt].
 \label{eq:noshedgas}
 \eeq
 The volume fraction enters  because the fluid occupies a smaller volume in the presence of gas so that the turbulence energy per unit  volume  decreases.
 The  dissipation length scale is smaller than the integral scale of the single phase flow,  so that $l_h < l$ in general.  
 Then for gas bubbles of {\em any size} that are not able to induce significant vortex shedding, one obtains turbulence suppression in general.
 
 The same type of analysis   holds for sufficiently small solids where vortex shedding does not occur even for appreciable turbulence levels. 
For large particle/fluid density ratio, $C_2\sqrt{k} \gg C_1$ and $C_2 \simeq  -2C_0\beta \f{\rho_p}{\rho_f}  \epdbar St/(1+St)$, so that
  \beqa
\f{k}{k_0} &\simeq& \f{l}{l_h} \lft[  1- \epdbar (1-\alpha \f{\rho_p}{\rho_f})  \rgt]  \non \\ &-&2\epdbar C_0\beta \f{\rho_p}{\rho_f}  \f{St \sqrt{k}}{(1+St)}\f{l_h}{C_\mu k_0}.
 \eeqa
 Here, $\alpha \rho_p/\rho_f \rightarrow 3/2$.
The meso-scale work  (last term) provides turbulence suppression as we have noted earlier.   In the  tracer limit,  $C_0 \rightarrow 0$, $\alpha \rho_p/\rho_f \rightarrow 1$, $l_h \rightarrow l$ 
and  $k=k_0$.  
 
 \subsection{The transition point}

Large amount of data \citep{GORE1989} indicate that the transition  between suppression and augmentation occurs  in a region around $D/l \simeq 0.1$. 
The transition point is given by $k=k_0$, and for small meso-scale work, it is given approximately by $k_{max} \simeq k_0$. From (\ref{eq:envelope}),
\beq
\epdbar (\alpha \f{\rho_p}{\rho_f}-1) \simeq   C_{\lambda}l/\lambda - \epdbar  B_V l/C_\mu,
\label{eq:TP}
\eeq
and one obtains 
\beq
\f{D}{l} \simeq \f{C_{\lambda}(6/\pi)^{1/3}(\epdbar) ^{-2/3}}{\alpha \f{\rho_p}{\rho_f}-1 + B_V \f{l}{C_\mu}},
\eeq
where $C_{\lambda} \simeq  k_\lambda$ in the region of interest (here,  $D \gg D_0$).
It was not possible to prove mathematically  that $D/l$ should be near 0.1 in general.   
In fact, the data of \cite{GORE1989} indicate a "fuzzy" transition region of $D/l$ in the range of about 0.05 to 0.3.
The important factor  that determines the value of the transition point in our model is the rapid increase of the energy injected by vortex shedding above a certain critical particle Reynolds number.

 \section{Comparison to data}

The model results were compared in detail to the data of \cite{MANDO} for glass beads in a jet,   the data compiled by \cite{CROWE2000} mainly for glass beads in air flow in a vertical pipe, and  to the often cited model of \cite{CROWE2000}. The  turbulence modification in terms of percentage change in  turbulence intensity  is  $\sqrt{k/k_0 -1}\times 100\%$, and this was compared to the experimental data.  Finally, we discuss  results for particles, droplets and bubbles in liquid in horizontal flow where only intrinsic vortex shedding contributes. Unfortunately, we  did not find the needed data in the literature for these latter cases.

\subsection{Model parameters, tuning and input parameters}

All  model parameters are listed  in Table \ref{tab:param}. After extensive trial and error we found that some of the parameters  could be treated as constants, while  three  parameters had to be tuned according to the type of carrier fluid in order to obtain reasonable results. We label the latter "tuning parameters" and their values  are given in Table \ref{tab:tuning} for all cases.
The  values of the constants  are given in the first half of Table \ref{tab:statistics} (above the double line). The only exception is the jet, where two constants are replaced with counterparts that vary  along the jet axis $x$,  as shown.

The main physical input parameters  are the layer thickness $h$, velocity difference $\Delta U$ over the layer,  volume fraction $\epdbar$ of particles, as well as the fluid and particle properties. All  input parameters   are given in the lower half of Table \ref{tab:statistics} (below the double line), for all cases.

\subsection{Non-dimensional numbers}
\label{sec:nondim}
The main non-dimension number  $D/l$, the ratio  between the particle diameter and the integral length scale. 
The Stokes number $St(D)=\tau_p(D)/\tau(D)$ is a central quantity   that influences the meso-scale work and the vortex shedding coefficient $B_V$. The eddy turnover time at the particle scale is
\beq 
\tau(D)=\epsilon^{-1/3} D^{2/3}.
\eeq 
The particle/fluid density ratio ($\rho_p/\rho_f$) enters in the particle relaxation time $\tau_p$. 
The   particle Reynolds number $Re_p(D)$  determines the onset of  vortex shedding and incorporates the continuous phase viscosity. The dissipation length scale ratio $l_h(D)/l$ determines the degree of dissipation, and   
the ratio $D/D_0$ determines when turbulence modification ceases at small scales. The Kolmogorov scale  is 
\beq
l_K=(\mu_c/\rho_c)^{3/4}\epsilon^{-1/4},
\eeq
and the small-particle cutoff scale $D_0$   is   proportional to $l_K$. 
It is noted that both $\tau$ and $l_K$  are a function of the dissipation rate $\epsilon$ which again varies with the (modulated) turbulent kinetic energy.


\begin{table} 
\centering  
\begin{tabular}{|l|c|}  
\hline
      & Definition    \\
 \hline
$k_v$   &  Vortex shedding calibration factor \\
 \hline
$n$ &  Turbulence spectrum slip velocity filter exponent\\
  \hline
  $\kappa_0$ & Largest scale wave-number in turbulence spectrum \\
 \hline
$Re_c$   &  Vortex shedding onset Reynolds number \\
 \hline
$p$   &  Vortex shedding onset exponent \\
 \hline
 $D_0$ &  Small diameter passive tracer cutoff scale \\
  \hline
$k_\lambda$ & Intra-particle dissipation length scale factor\\
 \hline 
 $c$ & Turbulence integral length scale factor\\
 \hline
 $\tau_{ad}/\tau$  & Added mass force and drag correlation time ratio \\
 \hline
\end{tabular}
\caption{\label{tab:param} Definition of the model  parameters.}
\end{table}	  

\begin{table*} 
\centering  
\scalebox{1.0}{

\begin{tabular}{|l|c|c|c|c|c|} 

\hline
    & Solid-Gas(pipe) & Solid-Gas(jet) & Solid-Liq.  & Drop-Liq. & Bubb.-Liq. \\
 \hline
$Re_c$   & 200     & 200      & 450    & 450    & 450   \\
 \hline
$k_\lambda$    & 0.2     & 0.2      & 0.05      &  0.05      &  0.05  \\
  \hline
$k_v$   & 6.0     & 6.1      & 6.0     & 6.0    & 6.0 \\ 

 \hline
\end{tabular}
 }
 
\caption{\label{tab:tuning} Tuning parameters.   The   parameters $Re_c$ and $ k_\lambda$ was found to be dependent on the carrier fluid; whether it is  gas or liquid. It is noted that $k_v$ is approximately the same for all cases.}
\end{table*} 

\begin{table*} 
\centering  
\scalebox{1.0}{
\begin{tabular}{|l|c|c|c|c|c|}  
\hline
    & Solid-Gas(pipe) & Solid-Gas(jet) & Solid-Liq.  & Drop-Liq. & Bubb.-Liq. \\
 \hline
$ p$   & 4     & 4      & 4    & 4    & 4    \\
 \hline
$c$   & 0.8     & $l(x)$      & 0.8      & 0.8     & 0.8      \\
 \hline
$\tau_{ad}/\tau_p$   & 0.1     & 0.1       & 0.1      & 0.1    & 0.1     \\
 \hline
$n$     & 0.3     & 0.3      & 0.3     & 0.3    & 0.3    \\
 \hline
$\kappa_0$   & $2\pi/h$ & $2\pi/(10 h(x))$ & $2\pi/h$ & $2\pi/h$ & $2\pi/h$    \\
 \hline
$D_0$   & $3 l_{K}$     &   $3 l_{K}$       & $3 l_{K}$     & $3 l_{K}$    & $3 l_{K}$    \\
 \hline
  \hline
$h$   & 5 cm (pipe radius)  & Increasing along jet & 5 cm  & 5 cm & 5 cm   \\
 \hline
$D$   &   10$\mu m - 8 mm$  & 1.3, 1.8 $mm$     & 1.2$\mu m - 15 mm$     & 1.0$\mu m - 11 mm$    & 1.4$\mu m - 13 mm$    \\
 \hline
$\epdbar$   & $ (0.05,0.5,1.6)\times 10^{-3}$     & $\epdbar(x) \in [10^{-5},10^{-3}]$      & $10^{-2}$    & $10^{-2}$    & $10^{-2}$   \\
 \hline
$\Delta U$   & 10 m/s   & $\Delta U(x)$ < 7.3 m/s      & 10 m/s     & 0.1-100 m/s   & 0.1-100 m/s    \\
 \hline
$\rho_c$   & 1.2 $kg/m^3 $     & 1.2 $kg/m^3$      & 1000 $kg/m^3$     & 1000 $kg/m^3$   & 1000 $kg/m^3$   \\
 \hline
$\rho_d$   & 2500 $kg/m^3$     & 2500 $kg/m^3$     & 1000 $kg/m^3$     & 800 $kg/m^3$    & 1.2 $kg/m^3$    \\
 \hline
$\mu_c$   & 0.018 cP     & 0.018 cP     & 1 cP    & 1 cP   & 1 cP   \\
 \hline
$\mu_d$   & -     & -      & -     & 1.25 cP    & 0.018 cP   \\
 \hline
$\sigma_{dc}$   & -     & -      & -     & 20 mN/m    & 20 mN/m \\  
 \hline
\end{tabular}
 }
\caption{ \label{tab:statistics} Model constants (above the double line). The numerical values were found to be the same for all cases, as shown.  The only exception is that the   integral length scale $l(x)$   increases along the jet axis $x$, and $c$ is not applicable.  The jet diameter is $2h(x)$, while the shear velocity $\Delta U(x)$ and $\epdbar(x)$ decreases as the jet spreads out. 
$D_0$ scales with  the Kolmogorov length $l_{K}$. 
The bottom half of the table below the double line  contains the material- and flow parameters for the different cases. The dynamic viscosity is measured in units of $1cP=10^{-3}$mPaS. $\sigma_{dc}$ is the interfacial tension between dispersed phase $d$ and continuous phase $c$. $h$ is the width of the domain or the shear layer, $D$ is the particle diameter and $\epdbar$ is the volume fraction of particles.  }
\end{table*}

\subsection{Glass particles in air flow: vertical pipe }

\begin{figure}
\centering
\includegraphics[width=1.0\linewidth]{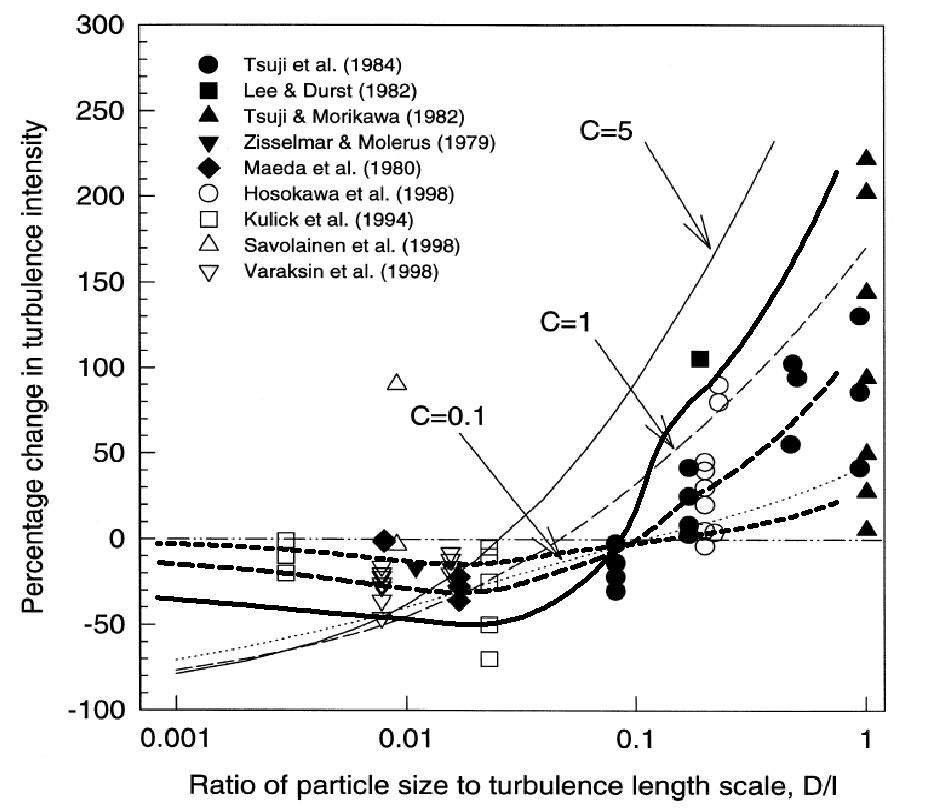}
\caption{The new model captures both the trends in  the turbulence suppression at small diameters, and turbulence enhancement above $D/l \simeq 0.1$ as shown by the thick lines for C=0.1 (dotted line), C=1 (dashed line), and C=5 (full line). Crowes model is represented by the thinner lines, but these increase monotonically with increasing diameter, in contrast to the new model.}
\label{fig:Crowe}
\end{figure}

 An often cited  model for turbulence modulation is that of \cite{CROWE2000}.
 A model example for a 10 m/s air flow with glass particles in a 10-cm pipe was compared in that work to data for similar experimental settings   (Figure \ref{fig:Crowe})\footnote{The cited work in the figure is referenced in \cite{CROWE2000}. }.  Crowe's model predicts a  monotonically  increasing trend for increasing diameter, as displayed for varying mass loading parameters  $C=\epd \rho_d/\rho_c =0.1,1.0,5.0$ (thin lines). However, the data-points show that  turbulence suppression  diminishes for smaller diameter, also in accordance with the data compilation by \citep{GORE1989}.  This trend was captured well with the new model as shown by the thick lines (with the same line styles for the different mass loading parameters). The  reason for this is the upgraded  dissipation model   now accounts for the passive tracer limit, whereas the dissipation length scale could be arbitrarily small in Crowe's model, leading to  overestimated dissipation for small $D/l$.    
 
\cite{CROWE2000} accounted for turbulence enhancement by vortex shedding due to gravitational settling of the glass particles, and terminal velocity ($V_\infty$) was adopted for the slip velocity.  Similarly, our model could fit the turbulence augmentation  for the larger diameters  only when this (extrinsic) source of turbulence was included (thick lines in Figure \ref{fig:Crowe} for the larger diameters).  The intrinsic vortex shedding source  in our model was  negligible  in comparison.  

The relaxation time $\tau_p$ for solids in gas is typically a few seconds, and most experimental setups for vertical flow with solids have a limited physical size that can very well imply a lower settling velocity in the measurement section  than the terminal velocity. 
A good fit to the data was obtained by setting
$U_{sg}= 0.63 V_\infty$,
corresponding to  a few meters per second for the millimeter sized glass particles.  A calibration factor of $k_{v}=6.0$   was used to get sufficient a sufficient level of turbulence augmentation. 
For the other model parameters, we adopted a critical Reynolds number of $Re_c=450 $ together with an exponent $p=4$, and $k_\lambda=0.2$.    
The remaining parameters  are found in Table \ref{tab:statistics} (first column).

\subsection{Glass particles in air flow: vertical jet}   
    Particle laden jets have been used extensively to obtain turbulence modulation data.  \cite{MANDO} performed experiments with a vertical air jet  laden with glass beads of diameter of 1-2 millimeter at mass loading ratios $Z$  (particle mass flux / gas mass flux) of the order of unity.  The particle to liquid density ratio was 2049.
    
\subsubsection{Experimental setup}   
 The author exploited the fact that $l$ increases downstream  along the jet such that $D/l$ decreases  for  a fixed particle diameter. The turbulence modulation was measured at different distances $x$ along the jet relative to the particle-free jet. A potential risk with this approach is that the flow conditions develop along the jet, in contrast to a steady, fully developed pipe or channel flow, and the  turbulence modulation at a specific $x$  may depend on the dynamics upstream  where $D/l$ is larger.  
 
 The inner nozzle diameter was 40 mm, and  the inlet air  velocity at the center   was   $7.3$ m/s.  The width of the jet increases linearly with distance $x$ from the nozzle, and we used the jet radius (at half the centerline velocity)    as  the   local  effective shear layer thickness  $h$ in the model. The centerline velocity in the jet diminishes with $x$, and we used this as an estimate of $\Delta U$ over the layer.   
Due to the spreading of the jet, the particle volume fraction  diminished from about  $10^{-3}$ close to the nozzle and to $10^{-5}$ at the maximum distance we considered downstream. The volume fraction was estimated as \citep{MANDO}
\beq
\epdbar=\frac{\dot{m}_p}{\rho_p U_p A},
\eeq
where $U_p$ is the particle centerline velocity, and $A$ is the jet cross sectional area. 
 The integral length scale $l$ increases with   $x$, and it was assumed that \citep{MANDO}, 
\beq
l=0.039 x\;[mm],
\eeq
based on  results   in the literature for particle-free jets. 
 We adopted  a scaling factor of $0.67 \times 0.039$, as a best fit value. A lower value  is reasonable as the particles introduce smaller length scales in the flow.

The particles lagged the air flow immediately after the nozzle by a few meters per second. The opposite situation was the case further downstream where the mean gas velocity was lower than the measured particle velocity. This behavior was due to gravitational settling. However, the corresponding particle Reynolds number was typically below 400, which indicates that vortex shedding was not necessarily very efficient. This is supported by the fact that turbulence enhancement would have been noticeable both at low and high $D/l$ corresponding to significant slip velocity near the nozzle and then further downstream, and this was not observed. 
Our model was then set up with the intrinsic vortex shedding effect only.     
 
\subsubsection{Comparison}  Data for varying mass loading and for two different glass  bead  diameters was compared to the model.
We tuned the model parameters to obtain an overall best fit and considered the three experimental  mass loading ratios Z=0.4, 0.95, 1.6.   A filter index of $n=0.3$ was used together with a calibration factor of $k_v=6.1$. 
The model is not very sensitive to   $n$ in this case, since the Stokes numbers are high in gas, and the filter response (of $W$ in equation \ref{eq:W}) is near unity. The  other   parameters   are found in Tables \ref{tab:tuning}  and  \ref{tab:statistics} (second column).
     
\begin{figure}[htp!]
\centering
\includegraphics[width=0.48\linewidth]{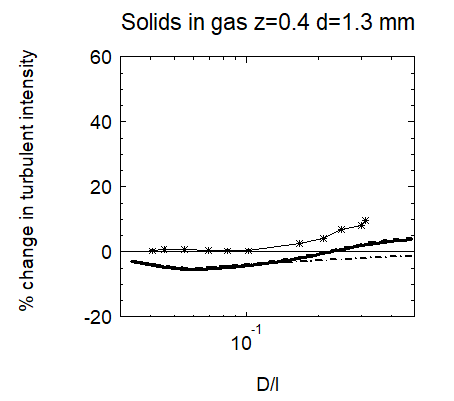}
\includegraphics[width=0.5\linewidth]{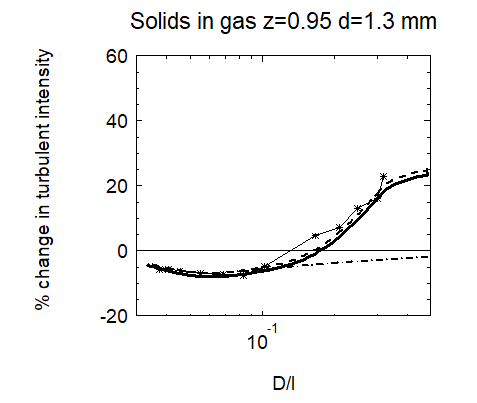}
\includegraphics[width=0.5\linewidth]{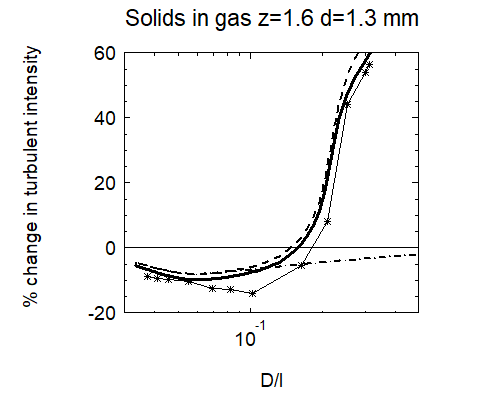}
\caption{Glass beads of diameter 1.3 mm in air jet for varying mass loading $Z$. The thick full line shows the model result. The   dashed line shows the result without the meso-scale work term (the maximum envelope \ref{eq:envelope}). The  dash-dot line shows the result with no vortex shedding, obtained  by setting $B_V=0$   (Equation \ref{eq:envelope2}).
}
\label{fig:MandoLeft}
\end{figure}

All three cases for $d=1.3$ mm are shown in Figure \ref{fig:MandoLeft}, and  the model reproduces the qualitative shape of the modulation profile  for all cases, but there is a systematic positive model offset   for Z=1.6 and and negative offset for Z=0.4, while the match is quite  good for the intermediate   case Z=0.95.  These results are  quite good, taken the uncertainties in the experimental data and the model approximations into consideration.  The data shows significant turbulence suppression at the   lower $D/l$ ratios for the cases Z=0.95 and 1.6, but practically no effect for Z=0.4 which may indicate an uncertainty  or bias in the dataset. 
Although not displayed here, it was found that the data for d=1.8 mm compared equally well to the model, using the same tuning parameters.

The model curves  in Figure \ref{fig:MandoLeft} 
demonstrate that  the  meso-scale work contributes very little as the full model (full line) is only marginally  below the maximum-envelope curve shown by the dashed line (representing equation \ref{eq:envelope}).   
 The turbulence level is then mainly controlled by a balance between production and particle-modulated dissipation. 
 Intrinsic vortex shedding sets in for  $D/l$ in the range 0.1-0.2 where the dash-dot line (equation \ref{eq:envelope2} with $B_V=0.0$) and the dashed line start to diverge.

\subsection{Neutrally buoyant solids in liquid} 

 \begin{figure} 
\centering
\includegraphics[width=1.0\linewidth]{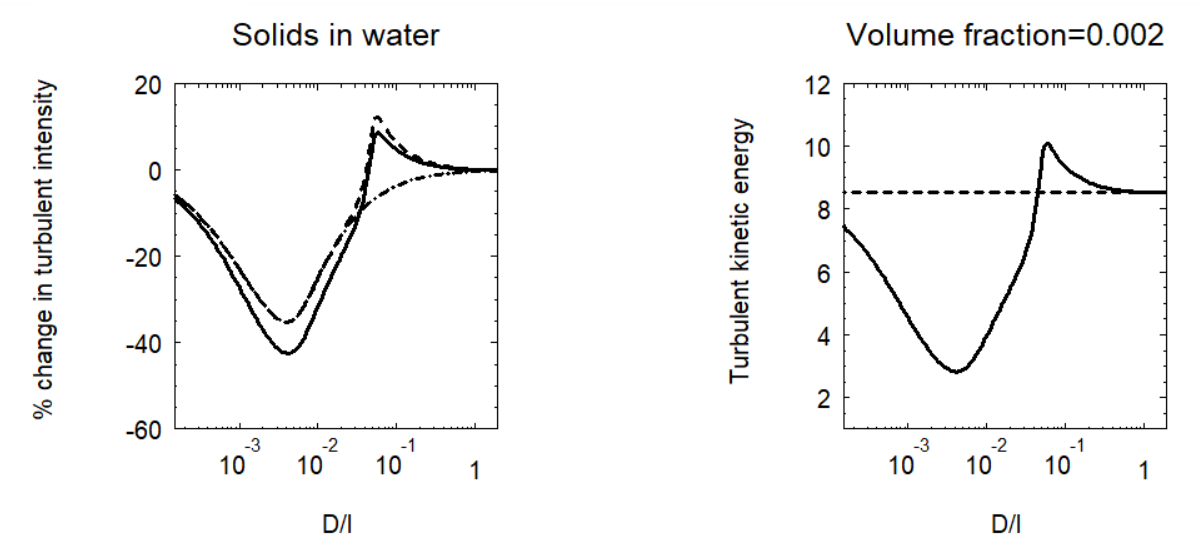}
\caption{Neutrally buoyant solids in water. 
  The full line shows the complete model. Right panel: The dashed line shows the result without the   meso-scale work term. The  dash-dot line shows the result with no vortex shedding (and no work). Left panel: The dashed line shows the unmodified turbulent kinetic energy per unit mass  $(m^2/s^2)$.}
\label{fig:solids}
\end{figure}

Neutrally buoyant particles in water with equal  particle and  liquid density  
is a desirable experimental case to avoid settling effects that lead to concentration gradients in horizontal flow and turbulence injection by gravitational settling in vertical flow.   

The  results  for solids in water with  volume fraction of  $10^{-2}$ are  shown  in Figure \ref{fig:solids}. The  parameters  are found in Tables \ref{tab:tuning} and \ref{tab:statistics} (third column).
The  full line shows the complete model (\ref{eq:master}). The   dashed line shows the result  without the meso-scale work term (\ref{eq:envelope2}). Again, the meso-scale work is negative, leading to   lowered turbulent kinetic energy.
The dash-dot line shows the result with no intrinsic vortex shedding.
The contribution from intrinsic vortex shedding is now   significant.  A low value of $n=0.3$ for the filter $W$  selects a wide range of wavenumbers, and the result is now   sensitive to $n$ since the Stokes number is lower in liquid mainly due to the lowered relaxation time $\tau_p$ relative to solids in gas. 
An important observation is that there  is less turbulence augmentation for larger diameter,   scaling as $1/D$ according to  (\ref{eq:vortexshed}),  since there are fewer particles when the volume fraction is held constant. This should  be a characteristic sign of intrinsic vortex shedding.  

The turbulence suppression for the smaller diameter range is now entirely due to the factor $l_h/l$ (seen by \ref{eq:envelope2} for $\alpha=1$), and $l_h$ decreases (dissipation increases) for smaller particle diameter as the particle separation decreases (the volume fraction is held constant). However, for very small particles near the Kolmogorov scale, the weighting with $C_\lambda$ is such that  $l_h/l\rightarrow 1$ and there is no turbulence suppression in the limit.     

\subsection{Droplets and gas bubbles in liquid} 

 Droplets and bubbles in turbulent liquid is a common situation in pipe flow.  The fundamental difference relative to solids, is that the size distribution is now set by the turbulence level rather than being a free parameter. Unless the droplet viscosity is large (enhancing the viscous work required for droplet breakup),  the maximum stable diameter is governed by the Hinze criterion for the competition between turbulence energy at the droplet size scale and the interfacial energy of the single droplet (interfacial tension). 
 First, a range of shear rates was invoked by varying  $\Delta U$, and a range of  turbulence length scales $l$ and dissipation rates  were calculated.     The Hinze criterion was used to calculate   a range of $d_{32}$ (Mean Sauter diameter) bubble or droplet sizes.  The dissipation rate  $\epsilon= C_{\mu}k^{3/2}/(\epbar l_h)$ was then updated with modified turbulence levels to calculate new sizes according to the Hinze-criterion. 
 
 Air bubbles in water with a particle to liquid density ratio of 0.0012, and  oil droplets  in water with an oil to water density ratio of 0.8 were chosen as test cases. A volume fraction of $2\times10^{-3}$ was chosen for both cases, and the same filter exponent of $n=0.3$ was used.
All model parameters are found in Tables \ref{tab:tuning} and \ref{tab:statistics}  (fourth and fifth columns). A drag coefficient of the same type as that of solids was assumed by resorting to contaminated interfaces of suppressed interfacial mobility. 

\begin{figure}
\centering
\includegraphics[width=1.0\linewidth]{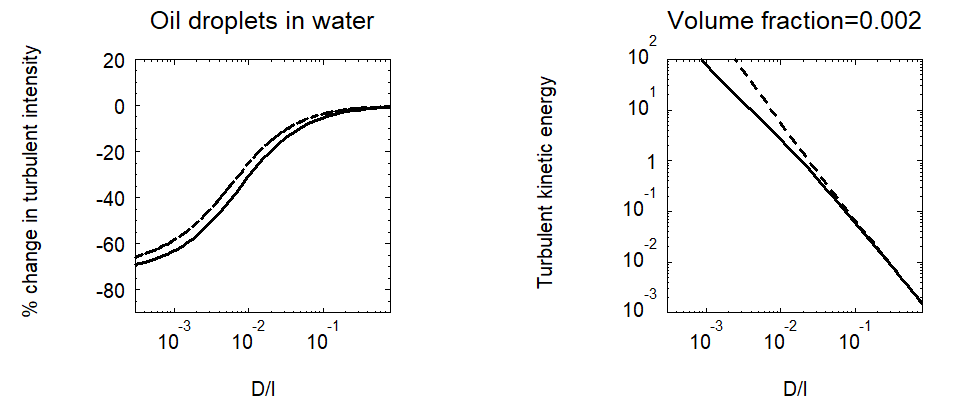}
\caption{Oil droplets in water for a volume fraction of $2\times10^{-3}$:  Left panel: The line coding is the same as in previous figures. Vortex shedding   does not contribute here since the turbulence kinetic energy is too small to generate an appreciable slip velocity, and the turbulence level is therefore not augmented.  Right panel: the dashed line shows the unmodified turbulent kinetic energy per unit mass  $(m^2/s^2)$ and the full line shows the modified turbulent kinetic energy.  }
\label{fig:drops}
\end{figure}

\begin{figure}
\centering
\includegraphics[width=1.0\linewidth]{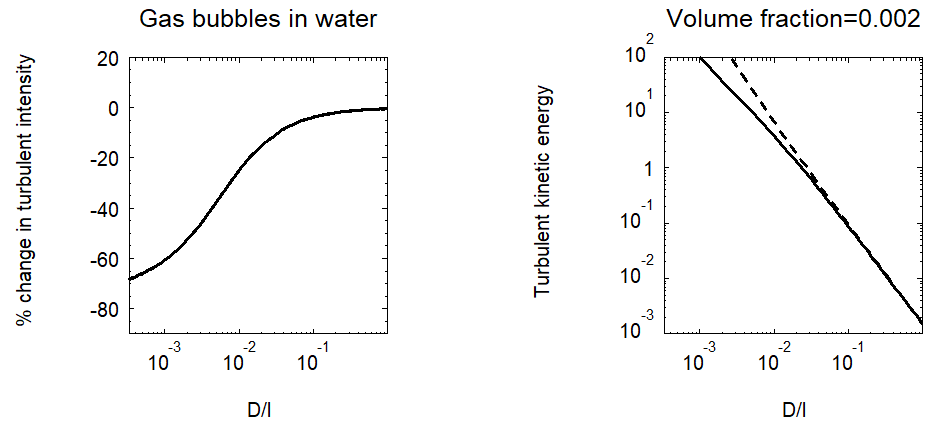}
\caption{Bubbles in water for a volume fraction of $2\times10^{-3}$.  Left panel:  The line coding is the same as in previous figures. Again, the vortex shedding source  does not contribute since the turbulence kinetic energy is  insufficient to cause an appreciable slip velocity. Right  panel: the dashed line shows the unmodified turbulent kinetic energy per unit mass  $(m^2/s^2)$ and the full line shows the modified turbulent kinetic energy. }
\label{fig:bubbles}
\end{figure}

The results are shown in Figure \ref{fig:drops} for droplets and  Figure \ref{fig:bubbles} for bubbles. 
 The   turbulent kinetic energy must be  smaller to support  larger drop/bubble sizes,  according to the Hinze criterion, and the  lowered turbulence level leads to  smaller slip velocities and less intrinsic turbulence injection by vortex shedding. Hence, turbulence augmentation   can be  negligible for droplets and bubbles, provided that  extrinsic vortex shedding  is also negligible.   This can be the situation in horizontal flow, but not necessarily in vertical flow where the settling velocity can build up to critical levels.  
For oil droplets, the meso-scale work plays some role (Figure \ref{fig:drops}) and  (\ref{eq:noshed}) is an appropriate model.  It is seen from Figure \ref{fig:bubbles} that neither  meso-scale work, nor intrinsic vortex shedding plays any role for gas bubbles at the  small volume fraction $2\times10^{-3}$, so that  (\ref{eq:noshedgas}) is a representative description. 

It is noted that the results for droplets and bubbles  are very similar, and this is due to the fact that it is now  the ratio $l_h/l$ that controls the level of turbulence suppression.
The turbulence suppression is   relatively large  (down to about$-60\%$) due to the small particle separation in this case (sizes are down to a few microns), and  this increases the dissipation rate considerably. A reduced factor $k_\lambda$  was adopted compared to the solids-in-gas cases to limit the dissipation rate. Lowered $k_\lambda$ corresponds to increasing the dissipation length scale relative to that in a gas, and this may be reasonable as the small scale inter-particle velocity gradients are expected to be smoother  for increased viscosity. 

  \begin{figure}
\centering
\includegraphics[width=1.0\linewidth]{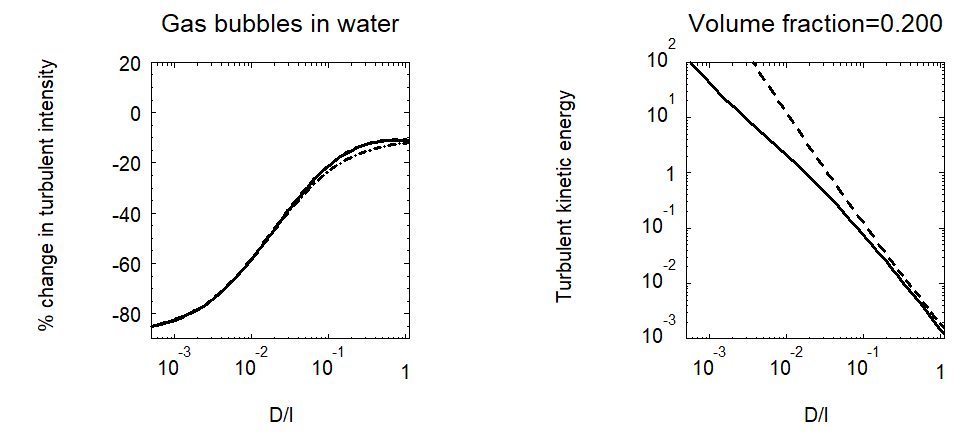}
\caption{Bubbles in water for a high volume fraction of $0.2$.    Left panel:  The coding is the same as in previous figures. The vortex shedding source to turbulence now contributes marginally at large diameter, but the turbulence level is now reduced for all diameters due to the large volume fraction (Equation \ref{eq:noshedgas}).  }
\label{fig:bubbles-high}
\end{figure} 

Higher volume fractions of gas bubbles is a common situation in pipe flow.  Bubbles in water for a  volume fraction of $0.2$ is shown in Figure \ref{fig:bubbles-high}.  The intrinsic vortex shedding source now contributes to a small degree at the larger diameters. The  slip velocity is small, but the volume fraction is larger by a factor of 100  to generate a small net vortex shedding contribution. Meso-scale work is again negligible, and (\ref{eq:noshedgas})  
is   representative. The high volume fraction of bubbles contributes with a lowering of the turbulent kinetic energy per volume of gas/fluid mixture via the factor $1-\epd$. Hence,  the turbulence level (per volume) is lowered correspondingly for all diameters, compared the case of smaller volume fraction in Figure \ref{fig:bubbles}.

\section{Discussion}

The large number of model parameters (given in Table \ref{tab:param}) reflects the complexity of the problem, as does the   extensive set of non-dimensional numbers that play an important role   (Section\ref{sec:nondim}).
However, most of the model parameters  could be taken as constants over   the cases studied. Only three parameters had to be tuned to the flow setting to obtain reasonable results; the critical Reynolds number for the onset of vortex shedding $Re_c$,  and  the intra-particle dissipation length scale factor $k_\lambda$  was set to vary according to the type of carrier fluid (liquid or gas). The scaling parameter of the vortex shedding source $k_v$ was nearly constant. 

A reduced factor $k_\lambda$ in liquid (relative to solids in gas) is reasonable, as smoother  small scale inter-particle velocity gradients would result with the increased carrier fluid viscosity. Increased critical Reynolds number for wake shedding is expected to occur for droplets or bubbles in liquid (relative to solids in gas)  since the  
interfacial mobility   delays  wake shedding for   lower particle to carrier fluid viscosity ratio \citep[][]{Clift1978}. 

The model is  sensitive to the critical  Reynolds number  $Re_p$, and the associated  exponent  $p$. The length scale $D_0$ below which the particles behave more like passive tracers   controls to a large degree the magnitude of the turbulence suppression in the lower diameter range. The results are not sensitive to the filtering exponent $n$ for large Stokes numbers (solids in gas), but they are more sensitive to $n$ for solids in liquid at lower Stokes numbers.  


The physical nature of the problem is highly nonlinear and iterations have to be used to obtain a consistent solution mainly due to the feedback from the  turbulence level  to the intrinsic particle slip velocity and the eddy turnover time. 
This non-linearity may lead to a divergent iteration if the intrinsic vortex shedding scaling factor ($k_v$) is too large.   
In contrast, the asymptotic forms (Sections \ref{sec:zerowork} and \ref{sec:zeroshed}) are simple algebraic formulae that can easily be implemented as reasonable approximations, without resorting to the full model. 

  
  We could not find sufficient data in the literature for bubbles, droplets, and neutrally buoyant solids in horizontal flow, and we encourage further experiments   for  horizontal liquid pipe or channel flow, starting with neutrally buoyant particles, and then moving on to droplets and bubbles. For the purpose of PIV (particle image velocimetry) for turbulence measurement, neutrally buoyant particles should be transparent, matching the index of refraction of the fluid to avoid light scattering effects.
  The use of modern x-ray technologies with suitable tracer particles may be used for opaque media or dense dispersions where PIV is not feasible. 
  
\section{Conclusion}
The model captures the trends in the data for solids in gas over  the full particle size range, with correct asymptotic behaviour of vanishing turbulence modulation  for small sizes due to the upgraded dissipation model.  
Model examples were also shown for solids, droplets and bubbles in liquid, with added mass forcing accounted for.   In general, there is  net work on the particles from the fluid at all diameters, contributing to a sink of  turbulent kinetic energy. This should be interpreted as removing energy from the fluid from the larger "meso-scale" length scales above the particle diameter. The energy gained by the particles is then dissipated through vortex shedding by injecting turbulence at the "micro-scale", and through particle boundary layer dissipation. 
This constitutes a second channel for the energy cascade to smaller scales, in addition to the usual breakup and transformation of vortex structures to smaller scales.

The negative meso-scale work 
is however a relatively small effect on the overall turbulence modulation compared to particle induced dissipation, vortex shedding and production due to large scale shear.
For very  small diameters, the particles behave  as passive tracers and there will be no   direct turbulence modification. For intermediate diameters, there is enhanced dissipation due to  induced velocity gradients in between
the particles and there is net turbulence suppression.
For   larger solids, slip between the particles and the fluid may cause significant  micro-scale turbulence injection by vortex shedding that is intrinsic to the turbulent flow. 
A good fraction of the data reported in the literature are for solids in vertical air flow, and we suspect that large  turbulence enhancement in some data (up to 300 percent)   is due to  gravitational settling and not intrinsic effects.

The transition point between turbulence suppression and augmentation occurs near $D/l=0.1-0.2$, and this seems to be a robust feature for a wide range of  experimental conditions. We could not prove  this to be  true in general, and it was  demonstrated that a transition is less likely to occur  for bubbles or droplets in liquid due to insufficient vortex shedding at the moderate turbulence levels that are needed to prevent breakup of the larger  bubbles and droplets. 
\newline  


%
%

%

\begin{acknowledgments}
This work was a closure for ideas sprung out from the Norwegian FACE research program in the period 2007-2014. A. Saber provided digitized data in the initial phase of the project. The work was funded internally  by IFE in 2022 via basic research funds.  
\end{acknowledgments}

\appendix

\section{Vortex shedding scaling factors}
\label{App:VSF}
The work on the fluid is distributed between viscous dissipation in the boundary layer of the particle, and turbulence energy in the wake. Thus, only a part of the work is converted to  turbulence, and we assume the redistribution factor 
\beq
f_{Re}=f_m(1-e^{-(Re_p/Re_c)^p}),
\eeq
such that the turbulence fraction approaches zero for small Reynolds number and all the work goes directly into viscous dissipation in the boundary layer. For high particle Reynolds number, we assume most of the energy goes into turbulence up to a certain fraction $f_m$. We chose $f_m=1$ for all cases studied.  $Re_c$ is a critical Reynolds number and $p$ is a tuning parameter.   
It is  necessary to provide a  rapid cutoff at a certain small particle size comparable to the Kolmogorov scale  to ensure no effect for the  smaller particles,  and  we incorporated the additional scaling factor
\bea
C_0=(1-e^{-(D/D_0)^2}).
\eea 
The scale $D_0$ was set to a few times the Kolmogorov scale.

\section{Phase averaging}
\label{App:Favre}

 If the relaxation time can be taken as a constant, the ensemble averaged force is 
\beq
\oln{\mathbf{F}} = \rho_p/\tau_p (\oln{\epd [\mathbf{v}]}-\oln{\epd [\mathbf{u}]}) -\rho_p\epdbar\mathbf{g}_e-\rho_p \alpha \oln{\epd \dot{\mathbf{u}}}
\eeq
To resolve the volume fraction-velocity products we used phase averaging (similar to Favre averaging   to derive the turbulence equations for compressible fluid) which is  defined as a combined ensemble and phase volume average, 
\beqa
{\widetilde A}_k=
\f{\oln{a^k[A]}}{\oln{a^k}}, \;\;\text{where} \; [A]=\f{1}{V_k}\int_{V_k} A dV
\eeqa
with averaging over the volume $V_k$ of phase $k$.
This approach may seem unnecessarily complicated, but it incorporates  the needed volume averaging over  the particle/fluid mix.
The components of the fluctuation velocity are then  defined relative to the phase averages $\uit$,
\bea
\uipp&=&u_i-\uit \non \\
\uit &\equiv & \f{\oln{a u_i}}{\oln{a}}. \non
\eea
Hence, the phase averaged velocity represents a "superficial velocity" or the volume flux of the phase normalised by the average volume fraction of the phase.  
  The Reynolds stresses and kinetic energy in (\ref{eq:phasic1D_k1}) are defined as 
\bea
\calS_{ij}  \equiv  - \f{\oln{\ep\uipp\ujpp}}{{\epbar}} \non \\
k= \f{1}{2}\f{\oln{\ep\uipp\uipp}}{{\epbar}}. \non
\eea

The ensemble averaged drag force can be written in terms of phase averaged quantities, and with $\beta=1/\tau_p$,   
\begin{widetext}
\beq
\oln{\mathbf{F}}_d =  \rho_p \beta  (\oln{\epd [\mathbf{v}]}-\oln{\epd(\mathbf{u}" + \widetilde{\mathbf{u}}}))  =\rho_p\beta \epdbar(\widetilde{\mathbf{v} } - \widetilde{\mathbf{u}}) - \rho_p\beta  \oln{(1-\ep)(\mathbf{u}" }) = \rho_p\beta \epdbar(\widetilde{\mathbf{v} } - \widetilde{\mathbf{u}}) - \rho_p\beta  \oln{\mathbf{u}" } .
\eeq
\end{widetext}

The last term is due to the difference between the volume average and the phase average, 
\beq
\oln{\mathbf{u}" } = \oln{[\mathbf{u}]} - \widetilde{\mathbf{u}} = -\f{\oln{\ep' \mathbf{u}'}}{\epbar}.  
\eeq
This represents a turbulent flux of fluid phase crossing the averaging volume, corresponding to an extra force on the fluid in the same direction. The fluctuating quantities $()'$   are here defined with respect to the straight ensemble average ($\mathbf{u}' = [\mathbf{u}] - \oln{[\mathbf{u}]}$, and $a'=a-\epbar$). 
 
Averaging the local fluid velocity over a   volume $V$ corresponds to an effective lowpass filtering of the energy spectrum. If we assume a standard (unmodified) Kolmogorov spectrum to leading order,   
\beq
\frac{u^2}{2} = C \epsilon^{2/3} \int_{\kappa_0}^{\kappa_L}  \kappa^{-5/3} d\kappa,
\eeq
where $\kappa_L=2\pi/V^{1/3}$ and $\kappa_0=2\pi/h$, where $h$ is the full extent of the flow domain.  The ratio $R$ between the filtered and unfiltered ($\kappa_L \rightarrow \infty$) variants is 
\beq
R=1-(L/h)^{2/3}.
\eeq
where the averaging length is $L=V^{1/3}$.
$R$ is close to unity if the averaging length $L$ is a few times the particle separation $\lambda$, and $\lambda \ll h$. In this case, most of the turbulent kinetic energy passes through the lowpass filter and the k-equation (\ref{eq:phasic1D_k1}) governs most of the energy contained in the spectrum.

\section{Dispersion tensor}
\label{app:lambda}
The dispersion tensor $\overline{\lambda}_{ji}$ is the correlation between the
local force ${f}_i({\bf x},t)$ on the particle in the i-direction, and the total displacement  $\Delta {\bf x}_j$ in the j-direction of the particle 
before it passes through ${\bf x}$ at time $t$,
\beqa
\overline{\lambda}_{ji}=\overline{ {f}_i({\bf x},t) \Delta x _j ({\bf x},t) }. \nonumber
\eeqa
These components are functions of the force correlation functions and Greens function of the EOM (\ref{eq:eom_simple}),
\beqa
\overline{ \lambda}_{ji}&=& \int_{-\infty}^{t}\overline{ {f}_i({\bf x},t){f}_k({\bf x}_p(s),s) } {G}_{kj}(t-s)ds,   \label{eq:lambda}  
 \eeqa
 where ${\bf x}_p(s)$ is the particle trajectory.
 Greens function $G_{kj}$ represents the particle displacement in the j-direction due to a force (delta function impulse) in the k-direction, and the total displacement in ${\bf x}_p$ is the time integral of $G_{kj}f_k$. The components $j\neq k$ is different from zero when there is a mean shear in the fluid. By considering drag and added mass, 
$f=d+a$, and 
$\overline{f_if_k}=\overline{d_id_k}+\overline{d_ia_k}+\overline{a_id_k}+\overline{a_ia_k}$. There will be four corresponding contributions
$\overline{ \lambda}_{ji}=(\overline{ \lambda}_{ji})_{dd}+(\overline{ \lambda}_{ji})_{da}+(\overline{ \lambda}_{ji})_{ad}+(\overline{ \lambda}_{ji})_{aa}$.
These components reduce to algebraic formulae in terms of the correlation timescales and local correlation tensors associated with the terms in $\overline{f_if_k}$, 
if we incorporate exponential correlation functions \citep{SKARTLIEN2009}.

\section{Iterative solution}
\label{App:iterate}
 
The "master equation" (\ref{eq:master})  can be recast to
\beqa
A\xi - B\xi^3 + C_1 \xi + C_2 \xi^2 = 0, 
\label{eq:master_app}
\eeqa
which is equivalent to the second order form  
\beqa
 B\xi^2 - C_2 \xi  - (C_1 + A) = 0,   
\eeqa
with formal solution
\beq
\xi_n=\frac{C_2 + \sqrt{C_2^2 +4B(C_1 + A)}}{2B}.
\eeq
As noted before, most of the coefficients   depend on $\xi$, leading to higher order algebraic relations that can be solved by iterations.
We handle this by letting the coefficients depend on  previous estimates $\xi_{\; n-1}$. The coefficients are 
\beqa
A = \zeta  \left[ \epbar + \epdbar \alpha \f{\rho_p}{\rho_f}  \right]  (\sqrt{2}/h) (\Delta U)^2, \\
B = C_\mu/l_h - \epdbar B_V(\xi_{\; n-1}), \\
C_1= C_0\beta\f{\rho_p}{\rho_f}  \lft[ S_b^1(\xi_{\; n-1}) 
+ \epdbar  S_f^1(\xi_{\; n-1})  \rgt] \label{eq:C1}, \label{eq:C_1} \\
C_2= C_0\beta \f{\rho_p}{\rho_f} \lft[ S_b^2(\xi_{\; n-1}) 
+ \epdbar  S_f^2(\xi_{\; n-1})  \rgt].
\label{eq:C2}
\eeqa
where the first and second order meso-scale work contributions are indicated by $C_1$ and $C_2$, respectively.
The dominating interaction term is  \newline $C_2 \simeq C_0\beta \f{\rho_p}{\rho_f}  \lft[ \epdbar  S_f^2  \rgt] = - 2C_0\beta \f{\rho_p}{\rho_f}  \epdbar St/(1+St) < 0$.

We encountered a problem with this formulation since $B$ can go negative during the iteration if the vortex shedding source is larger than  the  dissipation, and this leads to an imaginary root.
The workable approach is to incorporate the third order vortex shedding into the first order term with the following modifications 
\begin{widetext}
\beq 
C_1= C_0\beta \f{\rho_p}{\rho_f}  \lft[ S_b^1(\xi_{\; n-1}) 
+ \epdbar  S_f^1(\xi_{\; n-1})  \rgt] +  \epdbar B_V \xi_{\; n-1}^2  
\eeq
\beq
B=  C_\mu/l_h.
\eeq
\end{widetext}
It is confirmed that the converged solution satisfies the  "master equation" (\ref{eq:master}). 
For extrinsic vortex shedding,  
\beq
W_{vg}/\xi
\eeq
is added to the first order coefficient $A$ in    (\ref{eq:master_app}). Thus, $A \xi$  contributes with $W_{vg}$, which is  independent of the turbulence level and contributes as a constant in the algebraic equation.

\bibliography{bibfile}

\end{document}